\documentclass[aps,pre,showpacs,amsmath,twocolumn,a4paper]{revtex4}
\usepackage{graphicx}
\usepackage{verbatim}
\usepackage[pdftex]{hyperref}
\hypersetup{colorlinks=true,linkcolor=blue,citecolor=blue,urlcolor=blue}

\newcommand{\inth}{\int} %{\int_{-h/2}^{h/2}}
\newcommand{\inti}{\int_0^\infty}
\newcommand{\dby}[2]{\frac{\partial#1}{\partial#2}}
\newcommand{\var}{\operatorname{var}}
\newcommand{\cov}{\operatorname{cov}}
\newlength{\bwidth}
\newcommand{\bw}[1]{\makebox[\bwidth]{$#1$}}

\begin{document}

\title{Understanding the ideal glass transition: Lessons from an
  equilibrium study of hard disks in a channel}

\author{M. J. Godfrey}
\affiliation{School of Physics and Astronomy, University of
  Manchester, Manchester M13 9PL, UK}

\author{M. A. Moore}
\affiliation{School of Physics and Astronomy, University of
  Manchester, Manchester M13 9PL, UK}

\date{\today}
\begin{abstract}
We use an exact transfer-matrix approach to compute the equilibrium
properties of a system of hard disks of diameter $\sigma$ confined to
a two-dimensional channel of width $1.95\,\sigma$ at constant
longitudinal applied force.  At this channel width, which is
sufficient for next-nearest-neighbor disks to interact, the system is
known to have a great many jammed states.  Our calculations show that
the longitudinal force (pressure) extrapolates to infinity at a
well-defined packing fraction $\phi_K$ that is less than the maximum
possible $\phi_{\rm max}$, the latter corresponding to a buckled
crystal.  In this quasi-one-dimensional problem there is no question
of there being any \emph{real} divergence of the pressure at~$\phi_K$.
We give arguments that this avoided phase transition is a structural
feature -- the remnant in our narrow channel system of the hexatic to
crystal transition -- but that it has the phenomenology of the
(avoided) ideal glass transition.  We identify a length scale
$\tilde{\xi}_3$ as our equivalent of the penetration length for
amorphous order: In the channel system, it reaches a maximum value of
around $15\,\sigma$ at $\phi_K$, which is larger than the penetration
lengths that have been reported for three dimensional systems.  It is
argued that the $\alpha$-relaxation time would appear on extrapolation
to diverge in a Vogel--Fulcher manner as the packing fraction
approaches~$\phi_K$.
\end{abstract}

\pacs{05.20.--y, 64.70.Q--, 61.43.Fs}

\maketitle
\section{Introduction}
\label{sec:intro}

\begin{figure}
  \begin{center}
    \includegraphics[width = 3.5in]{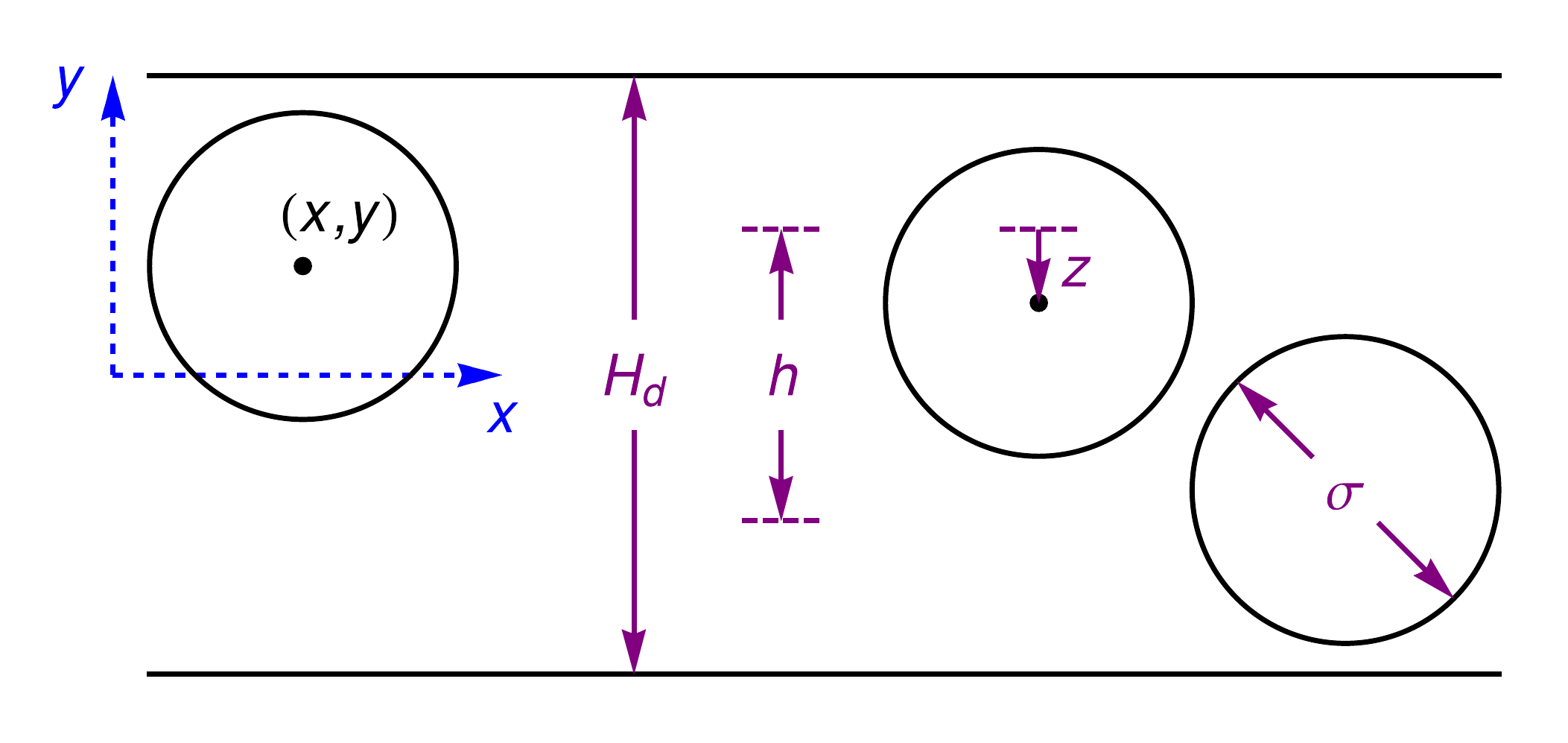}
    \caption{(Color online) Explanation of the notation used in this
      paper.  Here $H_d$ is the width of the channel, $\sigma$ is the
      diameter of each disk, and $h=H_d-\sigma$ is the width of the
      channel accessible to the centers of the disks.  We choose
      coordinates $(x,y)$ for each disk, where $y$ is measured from
      the center line of the channel.  The distance $z$ is the
      displacement of the center of a disk from the line of the
      largest possible displacement of a center from the middle of the
      channel, i.e., $z=h/2-y$.}
    \label{fig:notation}
  \end{center}
\end{figure}

There is a long history of studying exactly soluble one-dimensional
models in statistical physics.  While sometimes these models are
interesting in their own right, the main motive for their study is
usually to cast some light upon behavior in higher, more physical
dimensions $d$, such as $d=2$ and $d =3$, where exact solutions cannot
usually be found.  In this paper we study a quasi-one-dimensional
system of $N$ hard disks of diameter $\sigma$ confined by impenetrable
walls separated by distance $H_d=\sigma+h$, where $h$ is the width of
the region containing the centers of the disks (see
Fig.~\ref{fig:notation}).
For $\sqrt3\,\sigma/2<h<\sigma$, a disk may make contact with its
nearest- and next-nearest-neighbor (NNN) disks, but neighboring disks
cannot pass each other.  The ordering of the disks is still preserved:
$0\le x_1<x_2<\ldots<x_N\le L$, where $x_i$ is the position of the
center of disk $i$, measured along the channel, and $L$ is the total
length available to the disk centers.  The transverse coordinates of
the disk centers are denoted by $y_i$, where $|y_i|\le h/2$.  We have
in an earlier paper studied the case when $h<\sqrt3\,\sigma/2$, which
permits only nearest-neighbor (NN) contacts~\cite{Godfrey}.  All our
results and figures for the NNN model are for $H_d=1.95\,\sigma$.

\begin{figure}
  \begin{center}
    \includegraphics[width = 3.5in]{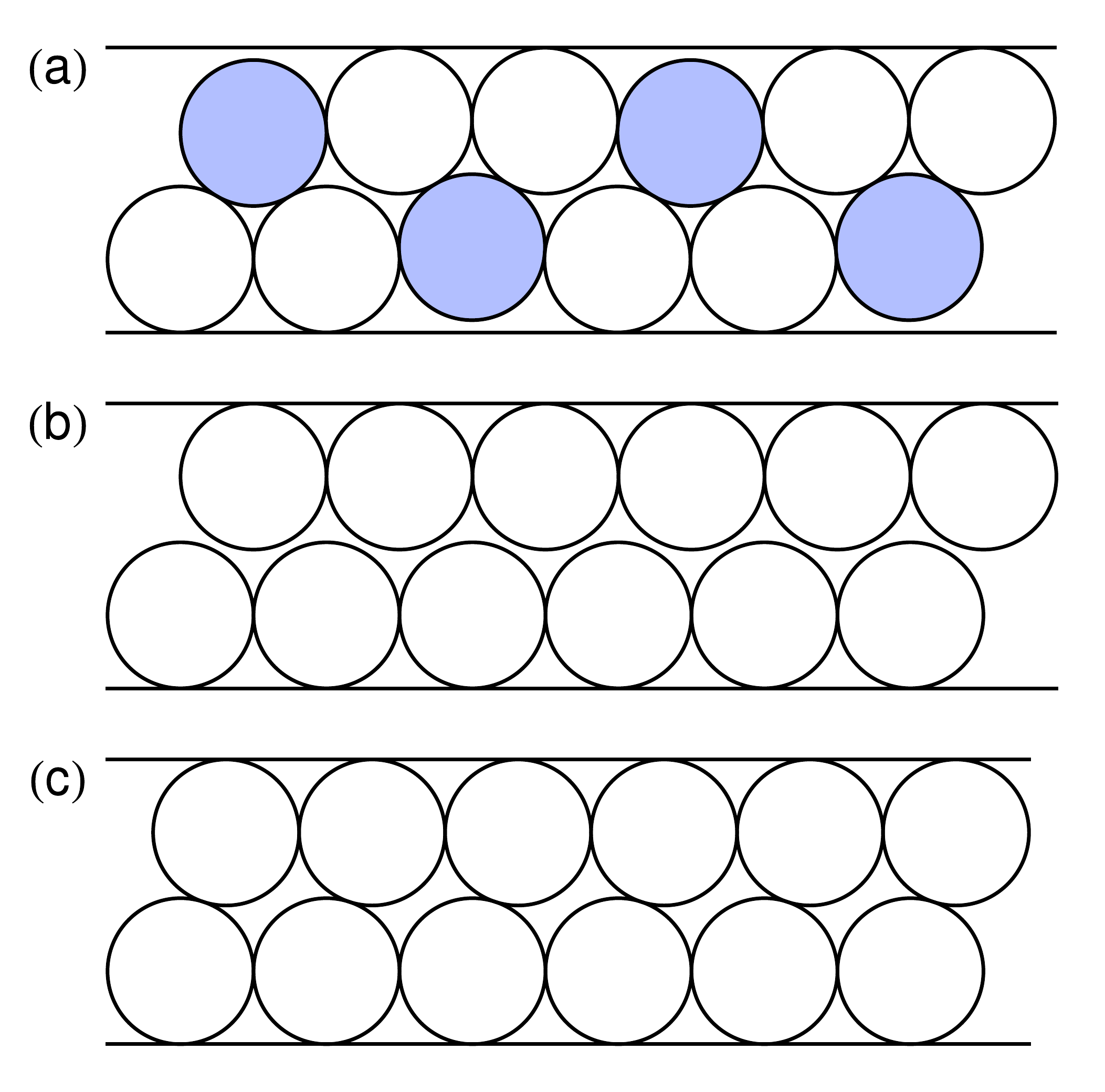}
    \caption{(Color online) (a) Configuration of disks at the maximum
      possible packing fraction $\phi_{\rm max} \simeq 0.8074$, when
      $H_d=1.95\,\sigma$.  The light blue disks do not touch the sides
      of the channel.  (b) Configuration of the disks at a density
      $\phi_K \simeq 0.8054$.  This is the highest-density state that
      can be reached if all disks are in contact with the channel
      walls.  It is not a jammed state as the disks in the upper row
      can be translated with respect to those in the lower row.  (c)
      Also a configuration at $\phi_K$ but at the limit of the
      translation of the upper row to the left.}
    \label{fig:buckled}
  \end{center}
\end{figure}

The numbering 1, 2, 3, \dots,~$N$ of the disks follows the ordering of
their $x$ coordinates.  In Fig.~\ref{fig:buckled}(a), the first disk
interacts with its nearest neighbor, the second disk, but it also
interacts with another disk, the third disk, giving rise to a NNN
coupling.  If instead we had $h<\sqrt3\,\sigma/2$, such NNN couplings
would not be possible: we call this case the \emph{NN model}.  Thus,
the NN and NNN models describe disks in narrow channels of different
widths.

For spin glasses, NNN interactions between the spins in a
one-dimensional system can introduce frustration.  Suppose we have
spins 1, 3, 5,~\dots\ in the lower of two rows of spins, and spins 2,
4, 6, \dots\ in the upper row, as shown below:
\begin{equation*}
  \begin{matrix}
      & 2 &   & 4 &   & 6 &   \\
    1 &   & 3 &   & 5 &   & 7
  \end{matrix}
\end{equation*}
With NNN interactions, spin 3 interacts with both of its nearest
neighbors 2 and~4, but also with its next-nearest neighbors 1 and~5.
This spin system has the same topology of interactions as will be
found for some arrangements of the disks in the narrow channel system.
While the spin system is effectively one dimensional, the spins in the
triangles 1--2--3 and 2--3--4 could be frustrated with appropriate
spin couplings.  Frustration is regarded as a key feature for
producing glassy behavior, so perhaps it is not surprising that the
presence of NNN contacts introduces features that are absent in the
narrower channel which has only NN contacts.

In the NN model, the dynamics becomes activated around a packing
fraction $\phi_d>0.48$~\cite{Godfrey,Ivan,Mahdi1,Mahdi2,Josh}.  This
is due to the onset of caging, a feature connected with the growth of
a particular structural feature, zigzag order~\cite{Godfrey,Josh}.
Zigzag order (see Eq.~\eqref{eq:zigzagdef} for its definition) can be
regarded as a form of bond-orientational order.  Accompanying this
activated dynamics are many of the other features which one associates
with glasses, such as the appearance of a plateau in time dependent
correlation functions, dynamic heterogeneities, etc.~\cite{Josh,Ivan}.
We expect there to be very similar behavior in the NNN model, but that
is not the focus of this paper.  Instead our interest is in a feature
which occurs \emph{only} in the NNN model, at a packing fraction
$\phi_K \simeq 0.8054$.  (The maximum packing fraction for $H_d
=1.95\,\sigma$ is $\phi_{\rm max}\simeq 0.8074$; see
Fig.~\ref{fig:buckled}(a).)  The phenomenology of this feature turns
out to be very similar to that of the ideal glass
transition~\cite{Zamponi}.

There are standard arguments against there being any genuine phase
transition in one dimension \cite{Cuesta} which we expect will apply
to our narrow channel problem, and will rule out any genuine phase
transition at finite pressure.  The ideal glass transition of hard
spheres in three dimensions is estimated to occur at a $\phi_K$ of
around $0.62$~\cite{Zamponi}; the analogue of the dynamical transition
at $\phi_d$ is around~$0.58$.  In two dimensions, activated dynamics
has been observed to set in at around $\phi_d \simeq 0.78$ in binary
\cite{Berthier} and weakly polydisperse \cite{Krauth} mixtures of hard
disks and the ideal glass transition is estimated to occur at a
packing fraction $\phi_K \simeq 0.81$~\cite{Zamponi}.  It is usually
believed that the dynamical transition at $\phi_d$ is avoided in any
finite dimension and it is uncertain whether the transition found at
$\phi_K$ in an approximate treatment, valid for dimension
$d\to\infty$, will survive to lower dimensions.  These approximate
calculations involve the analogue of one-step replica symmetry
breaking, which may be destroyed by fluctuation effects in any
physically interesting dimension~\cite{Moore,YeoMoore}.  Thus while
the absence of a genuine phase transition in our channel system is not
surprising, it is possible that glass transitions in two or three
dimensions might also be avoided.

A piece of evidence that the feature at $\phi_K$ might be the remnant
of the ideal glass transition is from the growth of a particular
correlation length, which grows to its maximum value (around
$15\,\sigma$) at $\phi_K$ and then falls.  We shall explain in
Sec.~\ref{sec:conclusion} that this length, which we call the
shear-penetration length $\xi_3$, is the distance over which a shear
displacement of the two rows of disks decreases with distance from an
amorphous boundary.  It is very similar in nature to the penetration
length for amorphous order calculated by molecular dynamics in
Refs.~\cite{Cavagnaetal,KobBerthier}.  The penetration length can be
defined in equilibrium (like our $\xi_3$) and is conceptually simpler
than the point-to-set length scale \cite{Cammarota,Cavagna}, which is
the length scale below which ergodicity is broken in a region of fluid
with amorphous boundaries; but both of these length scales are expected
to diverge at the glass transition~\cite{Biroli}.  In three
dimensions, the largest values found for the penetration length to
date are smaller than those we can find in our system, so in this
sense we have in our model better evidence for the ideal glass
transition than has been found in three dimensions.

The second aspect of our feature at $\phi_K$ which makes us identify
it as an avoided ideal glass transition is that the
$\alpha$-relaxation time $\tau_{\alpha}$, which we estimate using the
ideas in Ref.~\cite{Godfrey}, will appear to diverge in a
Vogel--Fulcher manner,
\begin{equation}
  \tau \sim \exp\left(\frac{{\rm const.}}{\phi_K-\phi}\right),
\label{eq:VFform}
\end{equation}
which is of the form expected for behavior near the ideal glass
transition.  Because there is no real singularity at $\phi_K$ as the
transition is avoided, the time scales will not truly diverge there
and will in fact diverge only for $\phi \to \phi_{\rm max}$.

There is a long tradition in the theory of glasses of attributing
glass behavior to the growth of particular types of structural order,
such as icosahedral order in three
dimensions~\cite{Tarjus,Royall,Royall2}.  Recent work by Cubuk et
al.~\cite{Liu2014} on the flow of jammed and glassy systems under
stress has shown that regions that are susceptible to rearrangement
can be discovered by machine-learning methods that combine information
derived from several features of the local structure, such as the
radial distribution of particles and the bond angles.  Since our
narrow-channel system with NNN contacts is relatively simple, we are
able to identify the collective motions of the disks that lead to the
apparent ideal glass behaviors.  Thus the onset of slow dynamics near
$\phi_d$ we can identify with the growth of bond-orientational order.
The feature at $\phi_K$ which seems related to the ideal glass
transition is connected with the growth of crystal-like order.  We
shall discuss this matter in detail in Sec.~\ref{sec:conclusion}.  In
the limit of very wide channels, when a real transition can arise
rather than just an avoided transition, the feature at $\phi_d$
becomes, we suspect, the fluid--hexatic transition, and the feature at
$\phi_K$ evolves to be the hexatic--crystal transition.  The length
scale $\xi_3$ appears to be similar to the penetration length for
amorphous order studied in the glass community.  For a system of disks
in a channel of any length, work must be done to displace one disk
along the channel, relative to its nearest neighbors.  But in a short
channel with length $L<\xi_3$, displacing one disk will cause the
displacement of all of the disks of one row relative to the other row,
as shown in Fig.~\ref{fig:buckled}(c).  This rigidity for $L<\xi_3$ is
like that of an amorphous solid.

Another example of the utility of the NNN channel problem for the
study of glasses is the work of Ashwin and Bowles~\cite{AshwinBowles}.
They determined exactly the number $N_c(\phi)$ of jammed states at
packing fraction $\phi$ and calculated from it the complexity
$S_c(\phi)=[\ln N_c(\phi)]/N$.  They found an apparent kink in
$S_c(\phi)$ at a packing fraction $0.8064$, a density which we shall
call $\phi_{\rm rcp} $, above which $S_c(\phi)$ decreases rapidly to
zero.  The jammed states which exist above this density contain
increasing amounts of the buckled-crystal ordering seen in
Fig.~\ref{fig:buckled}(a).  It is this feature which makes it natural
to identify the kink in $S_c(\phi)$ with random close packing, which
for hard spheres in three dimensions occurs at a packing fraction
close to $0.64$~\cite{Bernal}: jammed states with $\phi> 0.64$ are
known to contain increasing amounts of face-centered cubic crystal
ordering~\cite{Makse}.  Ashwin and Bowles also found that using the
Lubachevsky--Stillinger algorithm \cite{LS} to find jammed states
always led to a jammed state with a packing fraction close to
$\phi_{\rm rcp}$ if one started from a high enough initial density.
Observe that in our NNN channel system $\phi_{\rm max} \simeq 0.8074$,
$\phi_K \simeq 0.8054$, and $\phi_{\rm rcp} \simeq 0.8064$, which are
rather close to each other, compared to their corresponding values in
three dimensions.  These numerical values are for $H_d =1.95\,
\sigma$.  The general formulas are $\phi_K= \pi \sigma/(2H_d)$ and
$\phi_{\rm max}= 3\pi\sigma^2/ (2H_d\, a)$, where
\begin{equation}
  a =\sigma +2[\sigma^2 - (h-\sqrt{3}\,\sigma/2)^2]^{1/2}
\end{equation}
is the length of one unit cell of the buckled crystal; $\phi_{\rm
  rcp}$ has been estimated only for $H_d
=1.95\,\sigma$~\cite{AshwinBowles}.  Notice that $\phi_{\rm max}$
coincides with $\phi_K$ when $h=\sqrt3\,\sigma/2$ and that the
difference between them increases as $H_d$ increases.

From the results of simulations in three dimensions, it has been noted
\cite{Liu2007,Frenkel} that the rate at which distinct, disordered
states disappear with increasing density has a sharp maximum at a
particular density.  This maximum becomes narrower for larger system
sizes, perhaps becoming infinitely sharp and leading to a well-defined
$\phi_{\rm rcp}$ in the thermodynamic limit~\cite{Frenkel}.  Kamien
and Liu \cite{Liu2007} argue that such a feature, which corresponds to
a discontinuity in $S_c(\phi)$, will be accompanied by a divergence of
the pressure on a metastable branch of the equation of state.  They
suggest that the position of this singularity may be obtained by
extrapolation from the low-density portion of the equilibrium equation
of state.  We suspect that in our equilibrium calculation, the
apparent divergence of the force $F$, when extrapolated to $\phi_K$,
might have some connection with their argument.  It should be noted,
however, that the equilibrium state at $\phi_K$ is certainly not
jammed and that a snapshot of it would show much more disorder than in
Fig.~\ref{fig:buckled}(b), with disks at random distances from the
walls.  (This can be deduced from the density profile discussed later
in Sec.~\ref{sec:eqofstate} and illustrated in
Fig.~\ref{fig:density}.)  Using such a configuration as the starting
point for the Lubachevsky--Stillinger algorithm \cite{LS} will always
result in a slightly denser state, quite possibly one with packing
fraction close to~$\phi_{\rm rcp}$.  An interesting discussion of
possible singularities in the metastable fluid branch of the hard
sphere fluid is given in Ref.~\cite{Eisenberg}.

The plan of the paper is as follows.  In Sec.~\ref{sec:model} we set
up the transfer matrix formalism for the NNN channel problem.  It is
more intricate than that for the NN case and is harder to solve
numerically, especially for large packing fractions.  In
Sec.~\ref{sec:eqofstate} we study the equation of state and show that
the pressure apparently diverges when $\phi$ is extrapolated to
$\phi_K$.  We also obtain the local density variation across the
channel.  As the packing fraction increases, the disks are pushed more
and more against the channel walls, but for $\phi \sim \phi_K$ some of
the disks lift away from the walls in order to achieve the buckled
crystalline state of Fig.~\ref{fig:buckled}(a).  In
Sec.~\ref{sec:corrlengths} we obtain from the eigenvalues of the
transfer matrix three of the correlation lengths: the length scales
for zigzag and buckled-crystal order ($\xi_{zz}$ and $\xi_c$), and the
shear penetration length~$\xi_3$.  In Sec.~\ref{sec:probdensities} we
describe the distribution functions for the longitudinal separation of
nearest- and next-nearest-neighbor disks.  From these we can obtain
the variances of these separations, which are used in
Sec.~\ref{sec:lengthscalesexplained} to understand the variation of
$\xi_3$ with $F$, the longitudinal force applied to the system.  In
Sec.~\ref{sec:highdensity} we use the transfer matrix formalism to
describe how the system evolves towards the crystalline state as
$\phi\to \phi_{\rm max}$.  Finally in Sec.~\ref{sec:conclusion} we
further discuss the shear penetration length $\xi_3$ and explain why
the $\alpha$-relaxation time might be expected to show an apparent
divergence at~$\phi_K$.

\section{Model and transfer integral equation}
\label{sec:model}

In this section we set up the transfer matrix formalism for disks in a
channel that is wide enough to allow contact between next-nearest
neighbors.  Because a disk cannot overlap its nearest neighbors, it is
convenient to reparametrize the configuration not in terms of
$\{x_i,y_i\}$ but rather in terms of horizontal separations $s_i$,
defined by
\begin{equation}
  s_1\equiv x_1\quad\hbox{and}\quad
  s_i\equiv x_i-x_{i-1}-\sigma_{i,i-1} \quad \hbox{for $i\ge2$}\,,
\end{equation}
where
\begin{equation}
  \sigma_{i,j}\equiv\left(\sigma^2-[y_i-y_j]^2\right){}^{1/2}.
\end{equation}
Notice that when the disks touch, $s_i=0$.

In terms of the variables $\{s_i,y_i\}$, the configuration integral
can be written
\begin{align}
  Q(&N,L) \nonumber\\
  &{}= \prod_{i=1}^N\inti ds_i\inth dy_i \nonumber\\
  &\qquad{}\times\theta\bigl(L-\textstyle\sum_{j=1}^Ns_j-\sum_{j=1}^{N-1}\sigma_{j,j+1}\displaystyle\bigr)
  \prod_{k=2}^{N-1}\Theta_k\,,
\label{eq:QNL}
\end{align}
where $\theta(x)$ is the Heaviside step function and
\begin{equation}
  \Theta_k=
  \theta(s_k+s_{k+1}+\sigma_{k,k+1}+\sigma_{k,k-1}-\sigma_{k-1,k+1})\,.
\end{equation}
In Eq.~\eqref{eq:QNL}, the variables $y_i$ are integrated over the
range~$[-h/2,h/2]\,$; the same limits will be assumed in later
expressions that involve $y$-integrations.  Note that the first step
function in \eqref{eq:QNL} imposes the constraint $x_N\le L$ and the
remaining step functions exclude configurations in which
next-nearest-neighbor disks overlap.  A variant of our model which is
periodic in the $y$-direction has been studied in Ref.~\cite{Posch}.

It is inconvenient to work in an ensemble in which the length $L$ is
constant, though this was the approach taken by Barker~\cite{Barker}.
Instead we follow Kofke and Post \cite{Kofke} in transforming to an
ensemble in which the longitudinal force, $F$, is constant.  This
amounts to taking the Laplace transform of \eqref{eq:QNL} with respect
to~$L$,
\begin{align}
  \hat Q(\beta F,&N) \nonumber\\
  &{} = \inti e^{-\beta FL}\,Q(L,N)\,dL \nonumber\\
  &{} =\frac1{\beta F}
  \prod_{i=1}^N\inti ds_i\,e^{-\beta Fs_i}\inth dy_i \nonumber\\
  &\qquad{}\times e^{-\beta F\sum_{j=1}^{N-1}\sigma_{j,j+1}}
  \prod_{k=2}^{N-1}\Theta_k\,,
\label{eq:QNF}
\end{align}
where $\beta = 1/k_BT$.  Equation~\eqref{eq:QNF} agrees with the
results of Kofke and Post~\cite{Kofke} and Varga et
al.~\cite{GurinBalloVarga} who considered only the case $h/\sigma \le
\sqrt3/2$ in developing their transfer matrix formalism: in that
special case, the step functions are all unity, because
$\sigma_{k,k+1}+\sigma_{k,k-1}\ge\sigma_{k-1,k+1}\,$; the
$s_i$-integrations can be completed analytically, giving
\begin{equation}
  \hat Q(\beta F,N) = \frac1{(\beta F)^{N+1}}
  \prod_{i=1}^N\inth dy_i\,e^{-\beta F\sum_{j=1}^{N-1}\sigma_{j,j+1}}\,.
  \label{eq:QNFsimp}
\end{equation}
In this paper, we shall be concerned mainly with the case
$h/\sigma>\sqrt3/2$, so that we cannot make use of the
simplification~\eqref{eq:QNFsimp}.

\subsection{Formulation of an integral equation}
\label{sec:integraleq}

In \eqref{eq:QNF}, the integrals over $\{s_i,y_i\}$ can be performed
one at a time.  We write
\begin{align}
  \hat Q&(\beta F,N) \nonumber\\
  &{}= \frac1{\beta F}\inti ds_N\inth dy_N\inth
  dy_{N-1}\,q_N(y_N,y_{N-1},s_N)\,,
\end{align}
where the function $q_N$ can be calculated iteratively, starting from
\begin{equation}
  q_2(y_2,y_1,s_2) = \frac1{\beta F}\,e^{-\beta F(s_2+\sigma_{2,1})}\,.
\label{eq:qtwo}
\end{equation}
Subsequent functions $q_k$ (with $k=3$ to~$N$) are given by
\begin{align}
  q_{k+1}&(y_{k+1},y_k,s_{k+1}) \nonumber\\
  &{}=
  e^{-\beta F(s_{k+1}+\sigma_{k+1,k})}\nonumber\\
  &\qquad{}\times\inti ds_k\inth dy_{k-1} \,\Theta_k
  \,q_k(y_k,y_{k-1},s_k)\,.
\label{eq:iterator}
\end{align}
For $k\gg1$, the form of the function $q_k$ approaches that of the
nodeless eigenfunction corresponding to the largest eigenvalue
$\lambda_1$ of the integral equation
\begin{align}
  \lambda_n\,u_n&(y_2,y_1,s_2) \nonumber\\
  ={} &e^{-\beta F(s_2+\sigma_{2,1})}\inti ds_1\inth dy_0\,
  \Theta_1\,u_n(y_1,y_0,s_1)\,,
\label {eq:integraleq}
\end{align}
so that
\begin{equation}
  q_k(y_k,y_{k-1},s_k) \approx A\lambda_1^{k}u_1(y_k,y_{k-1},s_k)\,,
\label{eq:qapprox}
\end{equation}
where $A$ is a constant, independent of~$k$.  Accordingly, the free
energy $\Phi(\beta F,N)$ derived from the partition function $\hat
Q(\beta F,N)\sim \lambda_1^{N}$ is given by
\begin{equation}
  \beta\Phi(\beta F,N) = -\ln \hat Q(\beta F,N) = -N\ln\lambda_1 + {\rm O}(1)\,.
\label{eq:Philambda}
\end{equation}
This free energy is essentially \cite{quibble} the Gibbs free energy
with respect to the longitudinal force $F$, so that
\begin{equation}
  L = \dby\Phi F = -\frac N{\beta\lambda_1}\dby{\lambda_1}F\,,
\label{eq:LPhi}
\end{equation}
but it is also (via the dependence on the channel width) of the
Helmholtz type with respect the transverse force $F_T$,
\begin{equation}
  F_T = -\dby\Phi h = \frac N{\beta\lambda_1}\dby{\lambda_1}h\,.
\end{equation}
We are aware of no simple, general relationship between $F$ and $F_T$;
in particular, the stress tensor cannot be assumed to be isotropic.

It is of interest to see how the integral equation of
Ref.~\cite{Kofke} arises from \eqref{eq:integraleq} when
$h/\sigma\le\sqrt3/2$.  For that case, the step function $\Theta_1$ in
\eqref{eq:integraleq} is always unity, so that the dependence on $s_2$
reduces to $u_n(y_2,y_1,s_2)=\exp[-\beta Fs_2]\,u_n(y_2,y_1,0)$.  By
using this form in \eqref{eq:integraleq} and integrating both sides
with respect to $y_1$, we obtain
\begin{equation}
  \beta F\lambda_n\,\phi_n(y_2) = \inth e^{-\beta F\sigma_{2,1}}\,\phi_n(y_1)\,dy_1\,,
\label{eq:integraleqKP}
\end{equation}
where $\phi_n(y_1)=\inth u_n(y_1,y_0,0)\,dy_0$.  The one-dimensional
integral equation \eqref{eq:integraleqKP} is of the form derived by
Kofke and Post~\cite{Kofke}, with eigenvalue~$\beta F\lambda_n$.

\subsection{Properties of the integral equation}
\label{sec:symmetry}

The integral equation \eqref{eq:integraleq} has the same form when
all the $y$-coordinates are reversed in sign.  Each solution
consequently has either even or odd parity, $u_n(-y_2,-y_1,s)=\pm
u_n(y_2,y_1,s)$, corresponding to the parity of the solutions of Kofke
and Post's integral equation for the case of nearest-neighbor
interactions.

By using the symmetry of the function $\Theta_1$ under interchange of
$(y_2,s_2)$ with $(y_0,s_1)$, it is straightforwardly shown that real
eigenfunctions $u_n$ and $u_m$ corresponding to different real
eigenvalues $\lambda_n$ and $\lambda_m$ are orthogonal, in the sense
that the bilinear form
\begin{align}
  B[u_n,u_m] &{}\equiv \inti ds\inth dy_1 \inth dy_2\,
  e^{\beta F(s+\sigma_{1,2})}\nonumber\\
  &\qquad{}\times u_n(y_1,y_2,s)\,u_m(y_2,y_1,s)
\label{eq:Bfunctional}
\end{align}
equals zero for $\lambda_n\ne\lambda_m$.  This integral always
converges because, from \eqref{eq:integraleq}, both eigenfunctions
decrease as $\exp[-\beta Fs]$ for large~$s$.
Equation~\eqref{eq:integraleq} can also have complex eigenvalues,
which occur in complex conjugate pairs: the corresponding
eigenfunctions $u_n$ and $u_{n+1}\equiv u_n^*$ satisfy
$B[u_n,u_{n+1}]=0$ and are also orthogonal to the real eigenfunctions.
These complex solutions are used later in Sec.~\ref{sec:highdensity},
in a calculation of the correlation length at high density.

The orthogonality relation can be used to determine the coefficient
$A$ in Eq.~\eqref{eq:qapprox}.  Assuming that the function $q_2$ can
be expanded in terms of the eigenfunctions $u_n$,
\begin{equation}
  q_2 = A\lambda_1^{2}u_1 + \sum_{n>1}c_nu_n\,,
\end{equation}
projection of both sides onto the eigenfunction $u_1$ gives the exact
relation
\begin{align}
  B[u_1&,q_2] \nonumber\\
  &=\frac1{\beta F}\inti ds\inth dy_1 \inth dy_2 \,u_1(y_1,y_2,s) \nonumber\\
  &=A{\lambda_1^2}\,B[u_1,u_1]\,,
\label{eq:A}
\end{align}
which will be of use later, in Sec.~\ref{sec:distribution}.

The eigenvalue problem can also be formulated variationally.  The
eigenvalues are stationary points of the functional
\begin{equation}
  \Lambda[u] = \frac{K[u,u]}{B[u,u]}\,,
\end{equation}
where
\begin{align}
  K[u,v] = &\inti ds_1 \inti ds_2 \inth dy_0 \inth dy_1 \inth dy_2\,\nonumber\\
  &\quad\times u(y_1,y_2,s_2)\, \Theta_1\, v(y_1,y_0,s_1)\,.
\label{eq:Kfunctional}
\end{align}
This stationary property of the functional $\Lambda$ can be verified
by evaluating the functional derivative $\delta\Lambda/\delta u$ at
$u=u_n$, making use of the eigenvalue equation \eqref{eq:integraleq}
satisfied by~$u_n$.

The variational formulation of the eigenvalue problem can be used to
obtain the length of the system, by using \eqref{eq:LPhi} in the form
\begin{equation}
  \beta L/N = -\dby{}{F}\ln\Lambda[u_1]\,.
\label{eq:lambdalen}
\end{equation}
The functional $\Lambda$ depends on $F$ explicitly via the factor
$\exp[\beta F(s+\sigma_{1,2})]$ and implicitly via the
eigenfunction~$u_1$.  Terms in \eqref{eq:lambdalen} that arise from
the $F$-dependence of $u_1$ vanish, owing to the variational property
of $\Lambda$, leaving
\begin{widetext}
\begin{equation}
  L/N = \inti ds\inth dy_1 \inth dy_2\,
  (s+\sigma_{1,2})\,e^{\beta F(s+\sigma_{1,2})} u_1(y_1,y_2,s)\,u_1(y_2,y_1,s)
  /B[u_1,u_1]\,,
\label{eq:LbyN}
\end{equation}
\end{widetext}
which is analogous to Eq.~(2.12) of Ref.~\cite{Kofke}.  The form of
the last expression strongly suggests that
\begin{align}
  \rho(y_1,y_2,s) \equiv{} &e^{\beta F(s+\sigma_{1,2})}
  u_1(y_1,y_2,s)\nonumber\\
  &\qquad\times u_1(y_2,y_1,s)/B[u_1,u_1]
\label{eq:rho}
\end{align}
is the equilibrium distribution function for the horizontal separation
and $y$-coordinates of a neighboring pair of disks.  This
identification will be verified directly in the following section.  In
a similar way, the variational principle can be used to derive a wall
contact theorem \cite{Lebowitz,Percus} for the transverse force,
\begin{equation}
  \beta F_T/N = \dby{}{h}\ln\Lambda[u_1]
  = \rho_1(h/2)\,,
\label{eq:wallcontact}
\end{equation}
where $\rho_1(y)$ is the probability density for a disk to have its
center at~$y$:
\begin{equation}
  \rho_1(y) = \inti ds\inth dy_2 \,\rho(y,y_2,s)\,.
\label{eq:rho1def}
\end{equation}
{}

\subsection{Distribution functions}
\label{sec:distribution}

The probability distribution $\rho_k(y_k,y_{k-1},s_k)$ for neighboring
disks $k-1$ and $k$ to be found at separation $s_k$ with
$y$-coordinates $y_{k-1}$ and $y_k$ can be derived systematically by
performing all of the integrals in \eqref{eq:QNF} except those for
$s_k$, $y_{k-1}$ and~$y_k$.  The integrals over variables $s_l$ and
$y_{l-1}$ for $l<k$ give a factor of $q_k(y_k,y_{k-1},s_k)$ and those
for $s_l$ and $y_l$ for $l>k$ give a factor of
$q_{N-k+2}(y_{k-1},y_k,s_k)$.  The normalized result for $\rho_k$ is
\begin{widetext}
\begin{equation}
  \rho_k(y_k,y_{k-1},s_k) =
  q_k(y_k,y_{k-1},s_k)\,q_{N-k+2}(y_{k-1},y_k,s_k)\,e^{\beta F(s_k+\sigma_{k,k-1})}/\hat
  Q(\beta F,N)\,.
\label{eq:rhok}
\end{equation}
For the special case $k=N$, use of \eqref{eq:qtwo} shows that
\eqref{eq:rhok} reduces correctly to $q_N(y_N,y_{N-1},s_N)/\beta F\hat
Q(\beta F,N)$, which is normalized to unity.

In the thermodynamic limit $N\gg1$, most of the disks $k$ are far from
both ends of the system, so that $k\gg1$ and $N-k\gg1$.  We can make
the approximation \eqref{eq:qapprox} for $q_k$, $q_{N-k+2}$, and
$q_N$, giving
\begin{equation}
  \rho_k \to
  \rho(y_k,y_{k-1},s_k) =
      {\beta FA}{\lambda_1^2}
      \frac{u_1(y_k,y_{k-1},s_k)\,u_1(y_{k-1},y_k,s_k)\,e^{\beta F(s_k+\sigma_{k,k-1})}}
           {\inti ds_N\inth dy_N\inth dy_{N-1}\,u_1(y_N,y_{N-1},s_N)}\,,
\end{equation}
which, by using the result \eqref{eq:A}, is equivalent to~\eqref{eq:rho}.
\end{widetext}

\subsection{Numerical solution of the integral equation}
\label{sec:solution}

We solve the integral equation \eqref{eq:integraleq} by
discretization, approximating the integrals by sums.  This converts
\eqref{eq:integraleq} into a matrix eigenvalue problem
\begin{equation}
  \lambda \mathbf{u} = \mathsf{M}\mathbf{u}\,,
\label{eq:matrixevu}
\end{equation}
in which the matrix $\mathsf{M}$ is dense and nonsymmetric.  For large
values of $F$, the eigenfunctions are large when $y_1$ and $y_2$ lie
within a distance $\epsilon$ of the walls $y=\pm h/2$, where
\begin{equation}
  \epsilon=h-\sqrt3\,\sigma/2\,.
\label{eq:epsdef}
\end{equation}
It is important to treat these regions accurately without increasing
the dimension of $\mathsf{M}$ unnecessarily.  To achieve this, we make
a change of variables
\begin{equation}
  y(t) = at+b\tanh[ct]\,,
\label{eq:ytransform}
\end{equation}
where $y(\pm1)=\pm h/2$ and the values of $t$ are uniformly spaced in
the interval~$[-1,1]$.  The positive parameters $a$, $b$, and $c$ are
chosen so that approximately two thirds of the values of $y$ given by
\eqref{eq:ytransform} lie within a distance $\epsilon$ of the walls
and are nearly uniformly spaced within these narrow regions.

For simplicity, the eigenfunctions $u_n(y_2,y_1,s)$ are tabulated at
equal intervals in~$s$.  For any specified values of $y_2$ and $y_1$,
there is a value of $s$ (call it $s_{\rm max}(y_2,y_1)$) beyond which
the eigenfunctions $u_n(y_2,y_1,s)$ decrease as $\exp[-\beta Fs]$.
The portion of the $s$ integration with $s>s_{\rm max}$ can be
completed analytically, so that there is no need to tabulate $u_n$
beyond this point; this helps to reduce the dimension of the
eigenvalue problem.  Elsewhere, the integrands are linearly
interpolated from their tabulated values.

Even with the $y$ transformation and the treatment of the $s$
integration discussed above, the matrix $\mathsf{M}$ in
Eq.~\eqref{eq:matrixevu} is still too large to be stored in computer
memory, so that the eigenvalue problem must be solved by an iterative
method.  (With 100 points for each of $y_1$, $y_2$, and $s$,
$\mathsf{M}$ is a $10^6 \times 10^6$ matrix.  With fewer points we
have found it hard to capture accurately the rapid variation of the
eigenfunctions near the wall, as is visible in
Fig.~\ref{fig:density}.)  Simple iteration of \eqref{eq:integraleq} is
adequate for finding the first two real eigenfunctions, provided the
process starts with a function of the appropriate parity.  When more
eigenfunctions are required, we find it convenient to use the ARPACK
subroutine library \cite{ARPACK}, which implements the iterative
method due to Arnoldi~\cite{ArnoldiSaad}.

To test our numerical methods, we have used them to calculate the
equation of state and the correlation length $\xi$ for a system of
disks in a narrow channel with $h=\sqrt3\,\sigma/2$.  For this case in
which NNN disks do not interact, these quantities are more easily
found by solving the one-dimensional integral
equation~\eqref{eq:integraleqKP}~\cite{Kofke,Godfrey}, which leads to
a much smaller (and symmetric) transfer matrix.  Results from the two
approaches are in excellent agreement and give us some confidence in
the reliability of our later results for the case~$h=0.95\,\sigma$.

\begin{figure}
  \begin{center}
    \includegraphics[width = 3.5in]{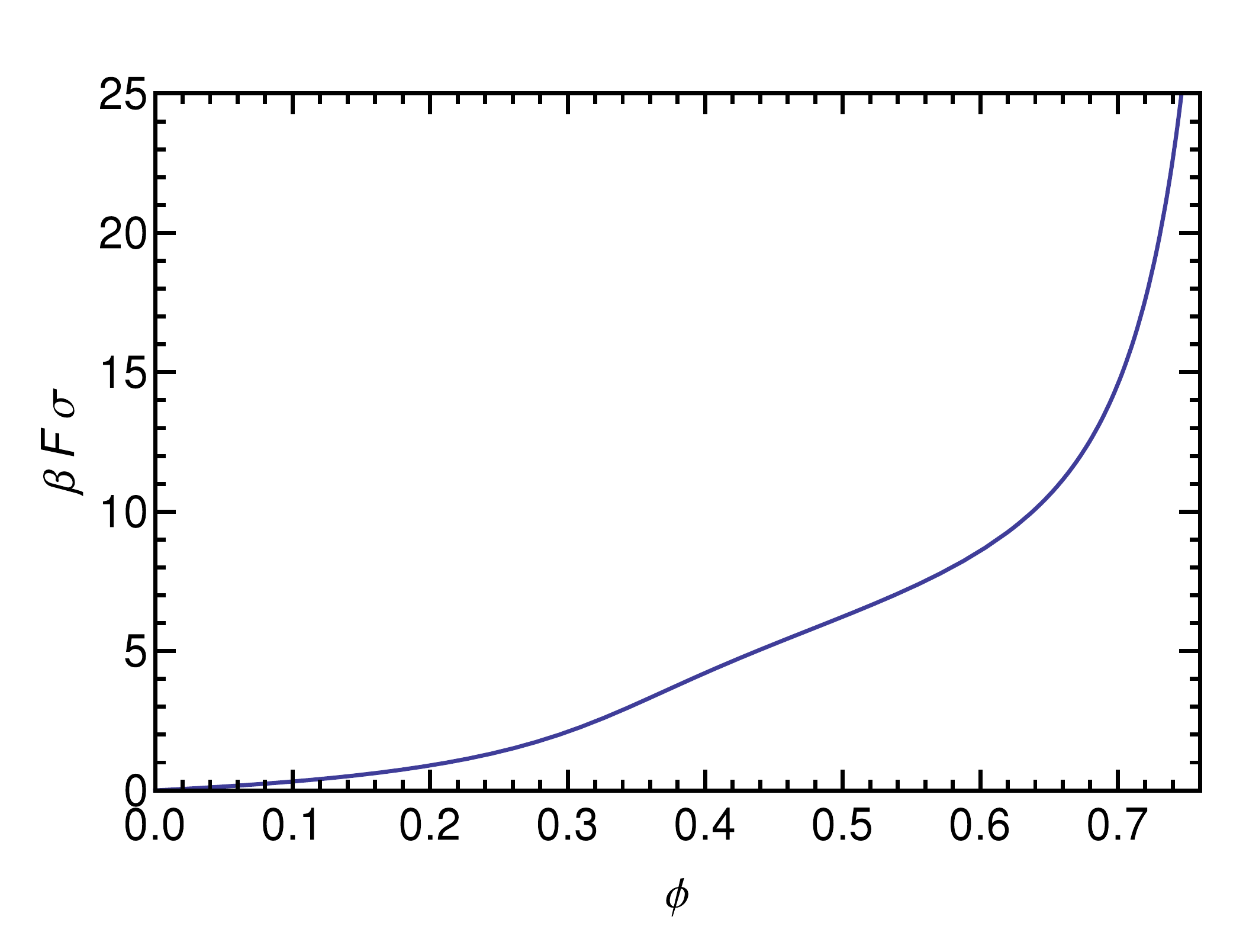}
    \caption{(Color online) Equation of state: $\beta F\sigma$ as a
      function of packing fraction $\phi$, for hard disks of radius
      $\sigma$ in a channel of width $H_d=1.95\,\sigma$.}
    \label{fig:eos}
  \end{center}
\end{figure}

\section{Equation of state}
\label{sec:eqofstate}

The equation of state is central to the study of fluids.  From it one
can determine the phase diagram.  Equation~\eqref{eq:LbyN} gives $L/N$
in terms of the $F$-dependent eigenfunction $u_1(y_2,y_1,s)$, and so
determines the equation of state; that is, the relation between $F$
and the packing fraction $\phi$, where
\begin{equation}
  \phi \equiv \frac{N\pi\sigma^2}{4LH_d}\,.
\end{equation}
The equation of state is shown in Fig.~\ref{fig:eos}, for moderate
values of~$F$ and $h=0.95\,\sigma$.  Notice the shoulder near
$\phi\simeq 0.5$: this is due to the onset of zigzag
order~\cite{Godfrey}.  If one could study progressively wider channels
we believe that this shoulder would evolve into the feature seen in
the equation of state of a two-dimensional fluid for $\phi\simeq0.71$,
which is the transition between the fluid and hexatic and crystal
transitions~\cite{KR}.  We expect that if one could study the
compressibility for a range of channel widths it would have a peak at
the shoulder which would evolve into the singularity seen in the
infinite system in Ref.~\cite{KB,KR}.  This is reminiscent of what
happens in (say) strips $n \times \infty$ of the two-dimensional Ising
ferromagnet: the peak in the specific heat for finite $n$ evolves into
a divergence at criticality as $n \to \infty$~\cite{YF}.

\begin{figure}
  \begin{center}
    \includegraphics[width = 3.5in]{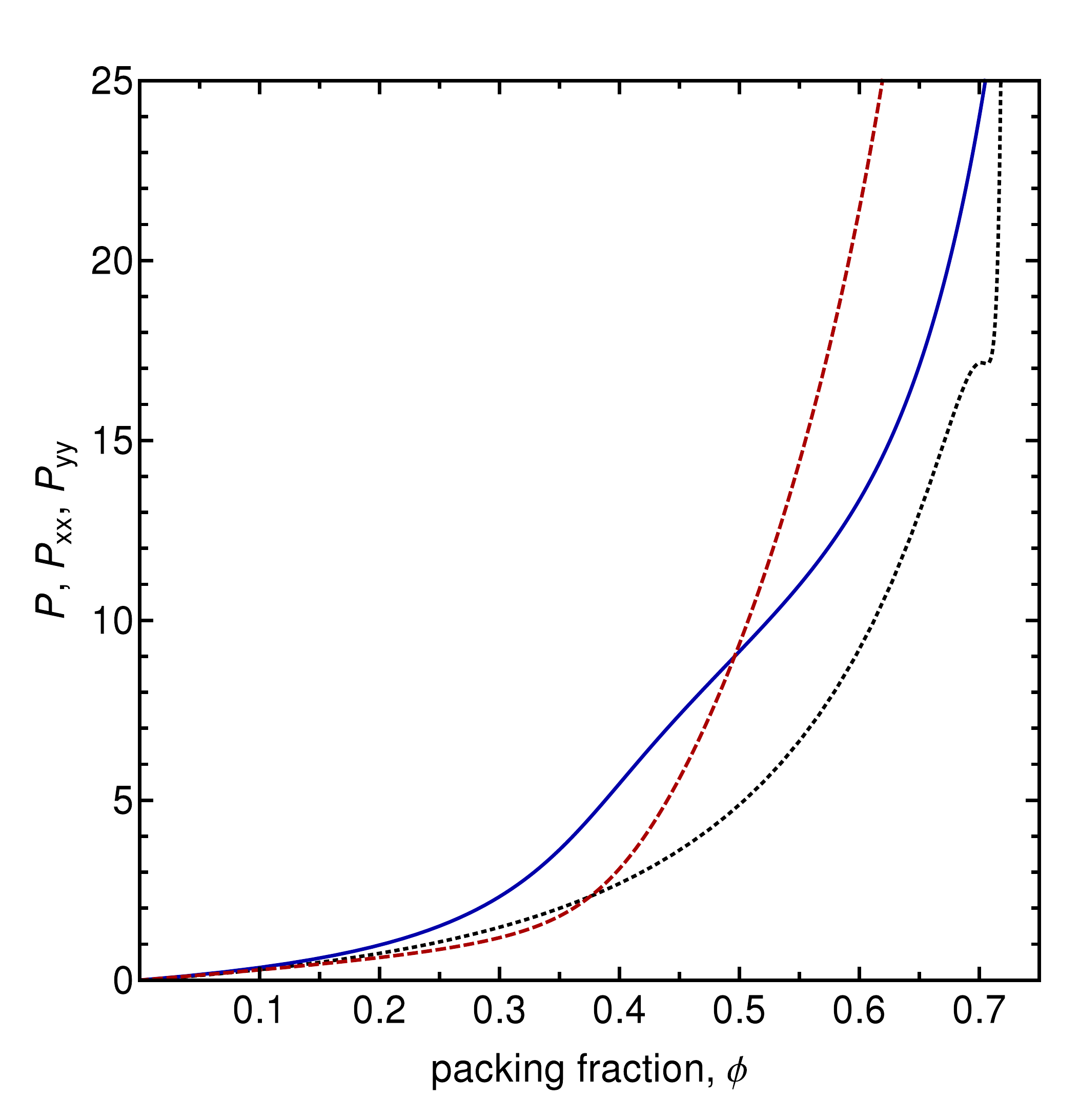}
    \caption{(Color online) Equation of state: $P$, scaled to $\beta P
      \sigma^2$, as a function of packing fraction $\phi$, for hard
      disks of radius $\sigma$ in a channel of width
      $h=\sqrt{3}\,\sigma/2$.  Similar results appear in
      Ref.~\cite{GurinBalloVarga}.  Here $P_{xx}$ is the longitudinal
      pressure $F/h$ (solid blue line), $P_{yy}$ is the transverse
      pressure $F_T/L$ (red dashed line), and the black dotted line is
      the simulation results of Kolafa and Rottner \cite{KR} on
      two-dimensional hard disks.}
    \label{fig:eosNN}
  \end{center}
\end{figure}

\begin{figure}
  \begin{center}
    \includegraphics[width = 3.5in]{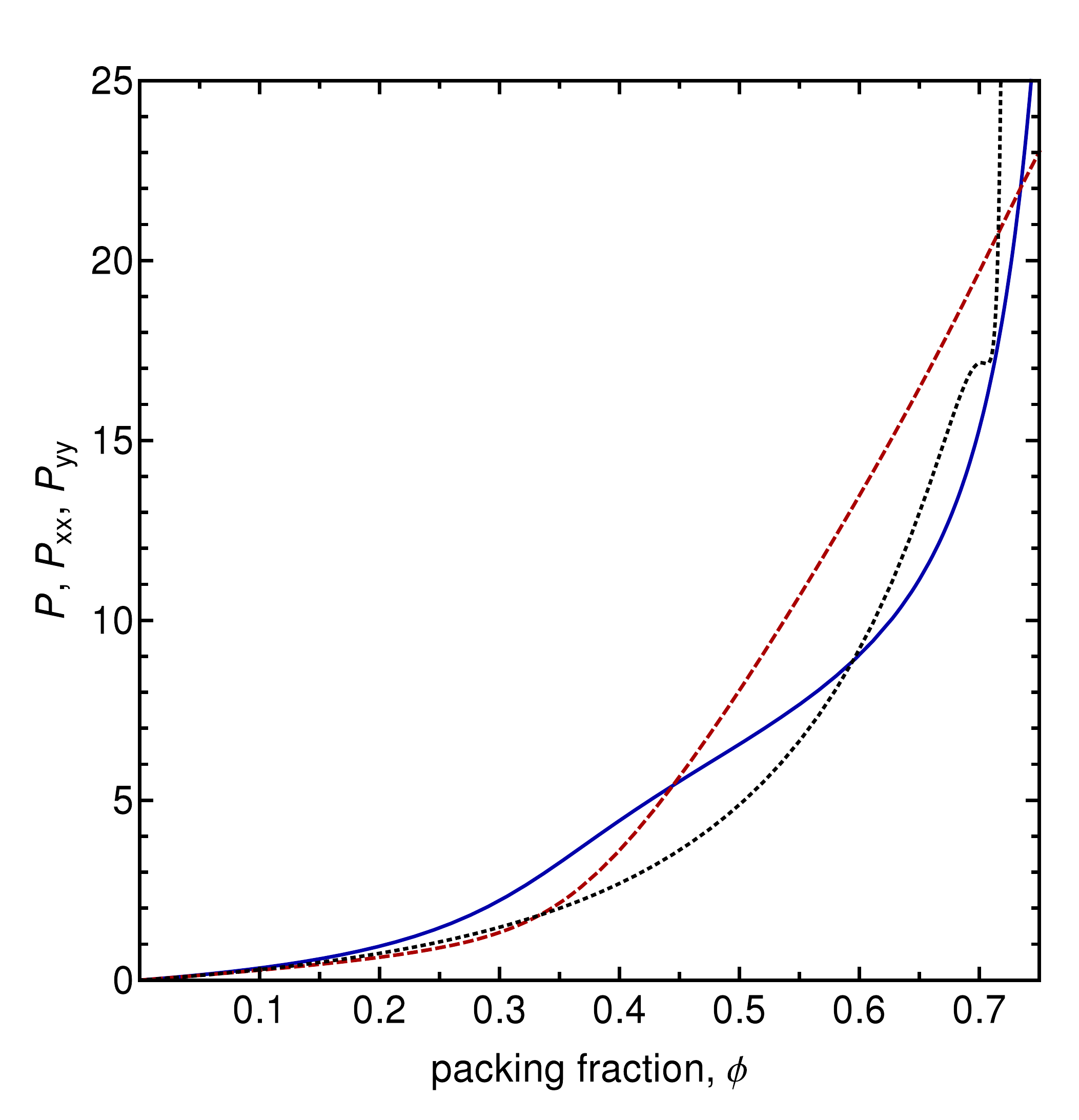}
    \caption{(Color online) Equation of state: $P$, scaled to $\beta P
      \sigma ^2$, as a function of packing fraction $\phi$, for hard
      disks of radius $\sigma$ in a channel with $h=0.95\,\sigma$.
      Here $P_{xx}$ is the longitudinal pressure $F/h$ (solid blue
      line), $P_{yy}$ is the transverse pressure $F_T/L$ (red dashed
      line), and the black dotted line is the molecular dynamics
      results of Kolafa and Rottner~\cite{KR}.}
    \label{fig:eos0.95}
  \end{center}
\end{figure}

In Figs.~\ref{fig:eosNN} and \ref{fig:eos0.95} we compare the
longitudinal and transverse pressures $P_{xx}=F/h$ and $P_{yy}=F_T/L$
calculated for the NN and NNN models with the equation of state of a
system of disks in two dimensions, simulated by molecular
dynamics~\cite{KR}.  Notice that at very low densities or packing
fractions, Figs.~\ref{fig:eosNN} and \ref{fig:eos0.95} show that the
equation of state approaches that of the two-dimensional system, which
is isotropic: $P_{xx}=P_{yy}$.  However, as the packing fraction
increases beyond about $0.1$, the system becomes sensitive to its
channel structure.  For wider channels, the approximate equality of
$P_{xx}$ and $P_{yy}$ for small $\phi$ should extend to higher packing
fractions, but probably not in any smooth way because of layering
effects.  It can be observed also that for the wider channel studied
in Fig.~\ref{fig:eos0.95}, the results are closer overall to those of
the full two-dimensional system than those for the narrower system
displayed in Fig.~\ref{fig:eosNN}.

\begin{figure}
  \begin{center}
    \includegraphics[width = 3.5in]{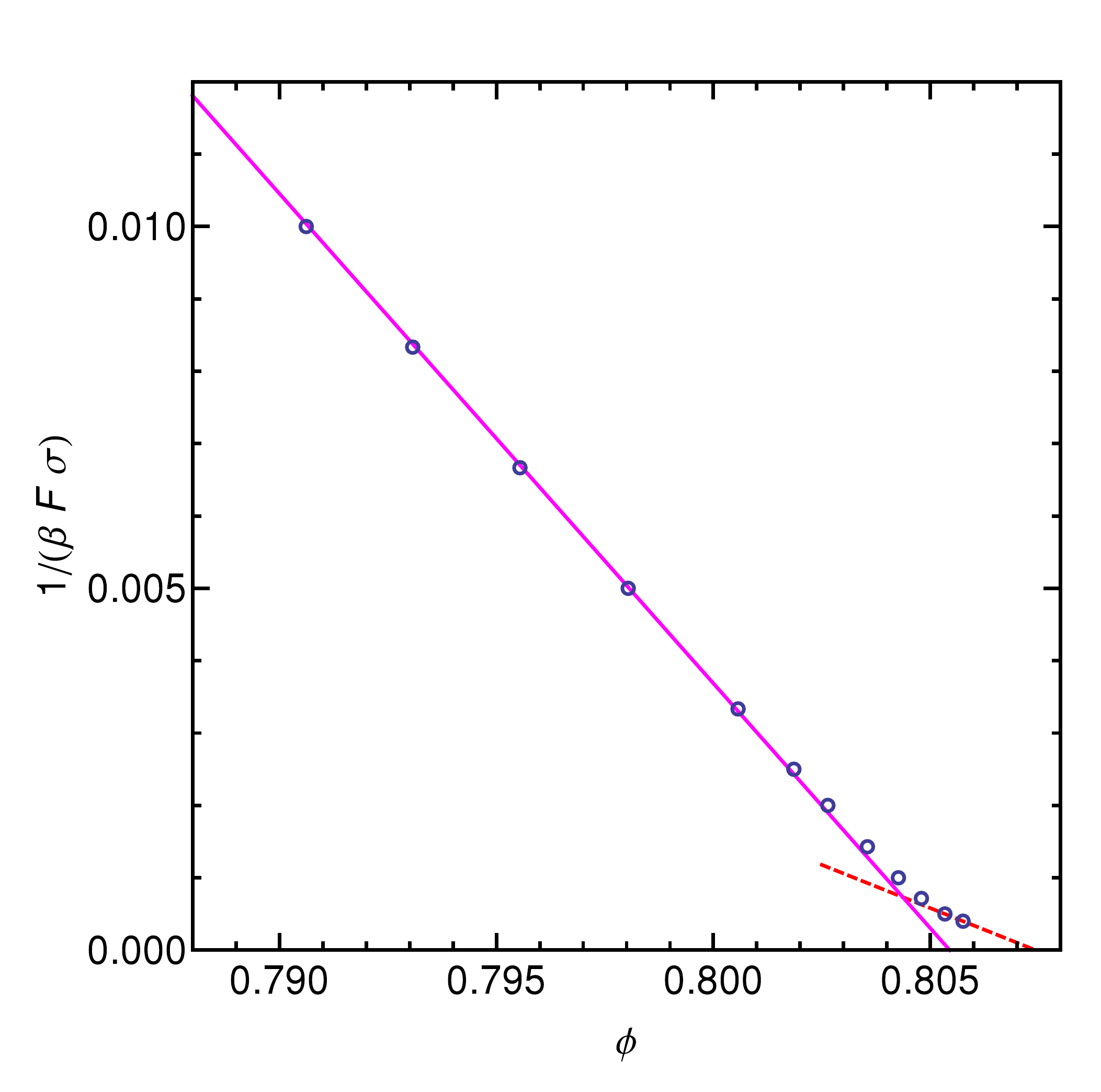}
    \caption{(Color online) Plot of $(\beta F\sigma)^{-1}$ against
      packing fraction $\phi$ for disks in a channel of width
      $H_d=1.95\,\sigma$.  The circles are our numerical results for
      $\beta F \sigma\le2500$ and the fitted straight solid line
      extrapolates to zero at $\phi=\phi_K=0.8054$.  The straight
      dashed line through the last two data points reaches zero at
      $\phi\simeq0.8074$, which is approximately $\phi_{\rm max}$.
      Its gradient corresponds to $d_{\rm ef{}f}=2.6$ rather than the
      value $2$ expected for $\phi\to\phi_{\rm max}$~\cite{SW}.  This
      suggests that the asymptotic regime has not yet been reached and
      that the plot will level off before approaching zero more
      steeply with a gradient corresponding to $d_{\rm ef{}f}=2$.  We
      have been unable to test this prediction by solving the transfer
      integral equation for~$\beta F \sigma>2500$.}
    \label{fig:onebyF}
  \end{center}
\end{figure}

The force $F$ becomes large for $\phi\to\phi_{\rm max}$, the density
of the most closely packed jammed state, which is the buckled crystal
identified in Ref.~\cite{AshwinBowles} and shown in
Fig.~\ref{fig:buckled}(a).  From the general arguments of Salsburg and
Wood~\cite{SW}, we might expect to find
\begin{equation}
  \beta F \simeq \frac{d_{\rm ef{}f}N}{L(1-\phi/\phi_{\rm max})}\,
\label{eq:SW}
\end{equation}
as $\phi$ approaches $\phi_{\rm max}$, with $d_{\rm ef{}f}=d =2$.
This form for the equation of state can also be obtained from the
high-density limit of the integral equation~\eqref{eq:integraleq}, as
discussed in Sec.~\ref{sec:highdensity}.  It is therefore of interest
to consider $1/F$ as a function of $\phi$, as shown in
Fig.~\ref{fig:onebyF}.  This shows a linear dependence on $\phi$ over
a range of $\phi$, extrapolating to zero at $\phi\simeq0.8054$, which
is \emph{not} the maximum density of the buckled crystal.  Instead,
the data in Fig.~\ref{fig:onebyF} is consistent with the equation
\begin{equation}
  \beta F \simeq \frac{d_{\rm ef{}f} N}{L(1-\phi/\phi_{K})},
\label{eq:fittoSW}
\end{equation}
with an effective dimensionality $d_{\rm ef{}f} \simeq 0.91$, at least
for the data that lies close to the straight line.  There is no actual
divergence of $F$ as $\phi \to \phi_K$, as the data points in the
vicinity of $\phi_K$ do not lie on the straight line.  The only true
divergence in $F$ is at the largest possible value of $\phi$,
$\phi_{\rm max} \simeq 0.8074$, which is the packing fraction of the
buckled crystal.  Our finding $d_{\rm ef{}f} \ne d$ has a counterpart
in the case of three-dimensional hard spheres~\cite{Maiti}, where
$d_{\rm ef{}f}\simeq2.53$ rather than~3.  Notice that our $d_{\rm
  ef{}f}$ is close to $1$.  This may be because in
Fig.~\ref{fig:buckled}(b) the top row and the bottom row are partially
decoupled so that each behaves like a one-dimensional hard rod gas
of maximum density~$\phi_K$.  Each row contains $N/2$ disks which as
$\phi \to \phi_K$ will exert a force $F_R$ as for hard rods: $ \beta
F_R=N/[2L(1-\phi/\phi_K)]$.  The combined force exerted by both rows is
$F=2 F_R$.

The packing fraction $\phi_K\simeq0.8054$ at which the force appears
to diverge corresponds to the configuration shown in
Fig.~\ref{fig:buckled}(b).  This is not a jammed state as it can be
deformed into Fig.~\ref{fig:buckled}(c) by sliding the top row of
disks to the left.  As the density increases above $\phi_K$, disks
have to move off the wall to fit into the system and the buckled
crystal state starts to develop.  At all but the highest forces,
however, the shaded disks in Fig.~\ref{fig:buckled}(a) are only
lightly pinched by their neighbors, so that the entropy gained by
allowing these disks to be delocalized near the walls overcomes the
work that must be done to lengthen the system.

\begin{figure*}
  \begin{center}
    \includegraphics[width = 3.5in]{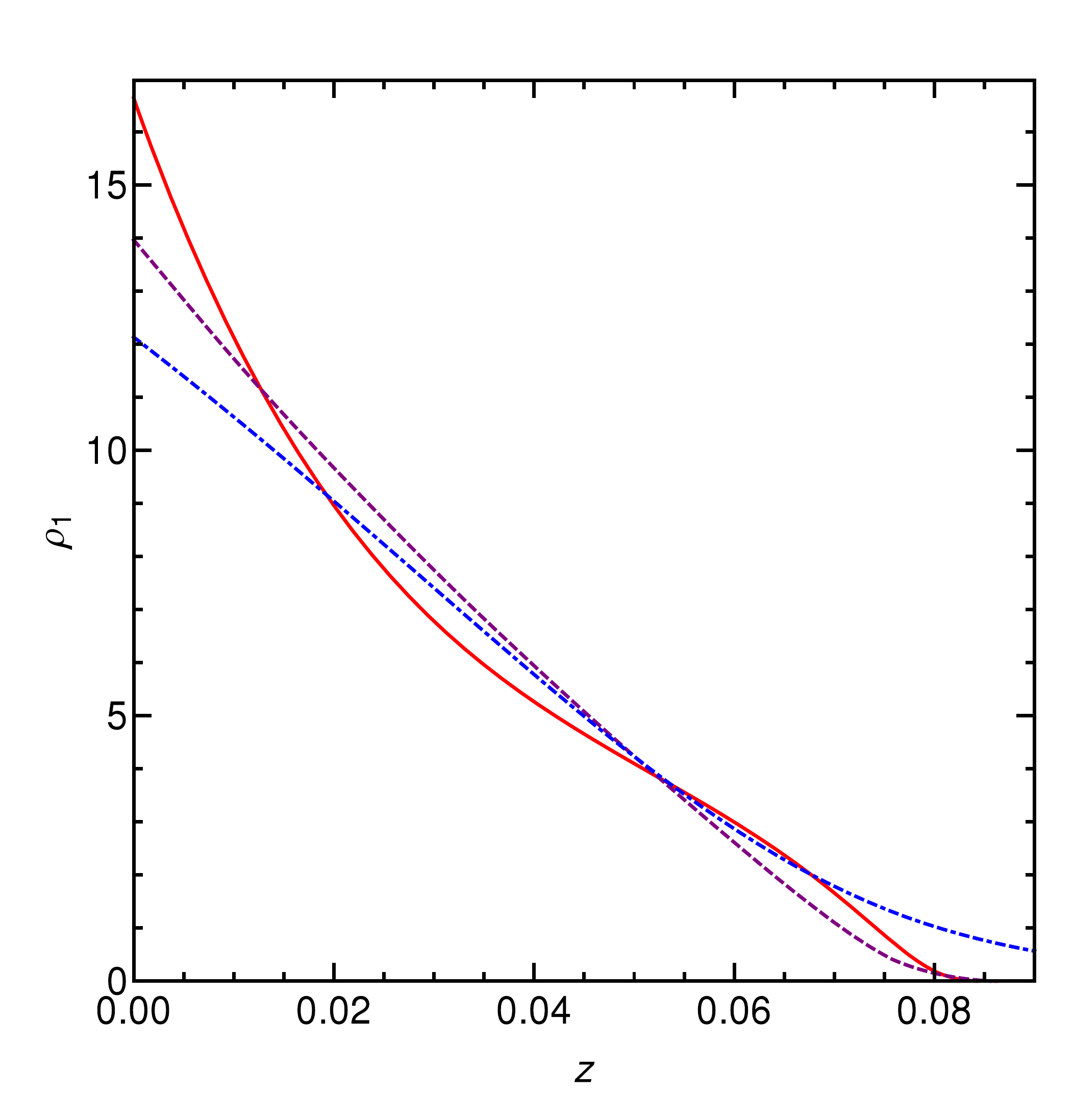}
    \includegraphics[width = 3.5in]{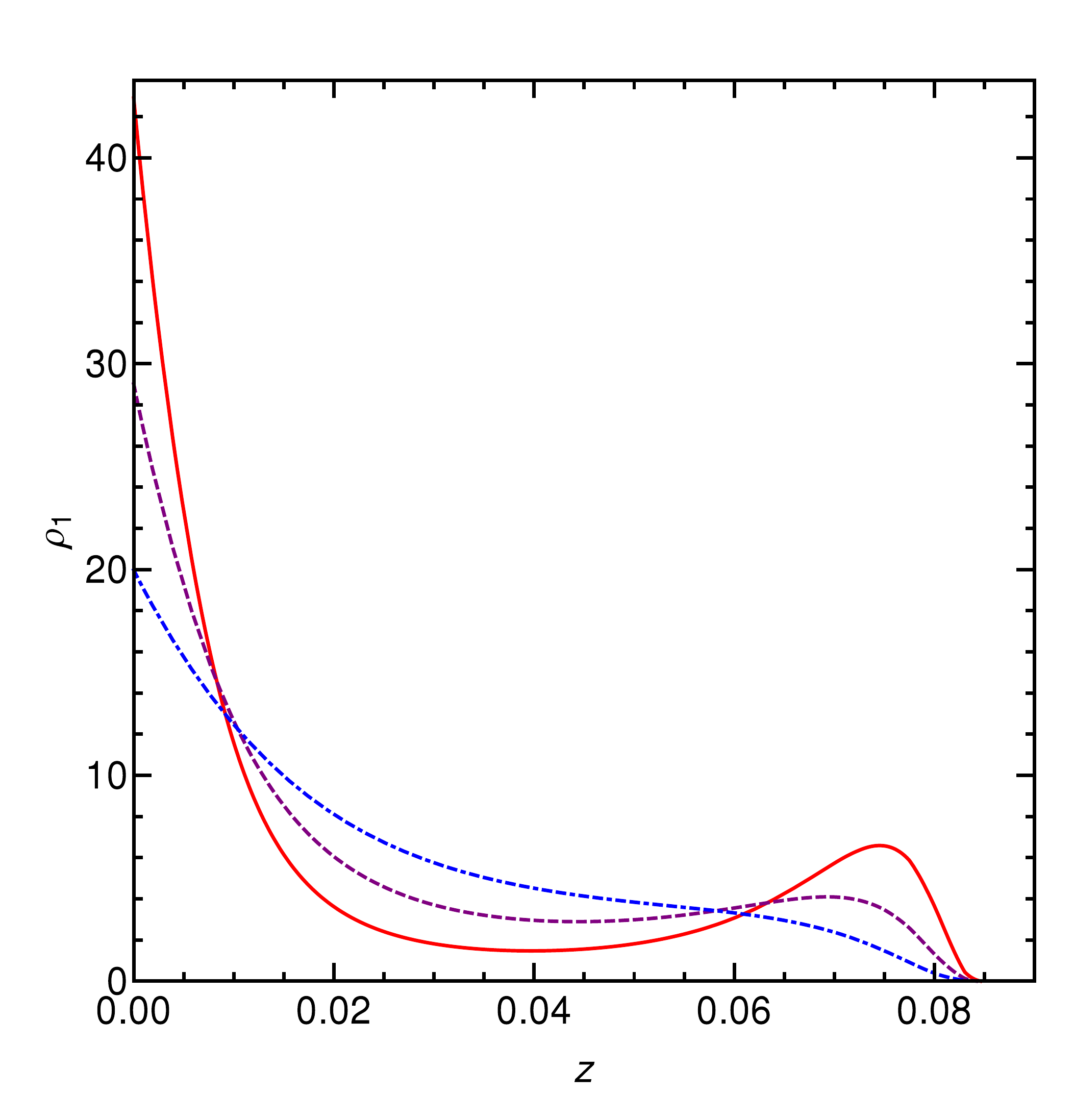}
    \caption{(Color online) Distribution function
      $\rho_1\bigl(\frac12h-z\bigr)$ for the position of a disk in a
      channel with $h=0.95\,\sigma$; $z$ is the distance (in units of
      $\sigma$) from one wall of the channel containing the disk
      centers.  On the left the dash-dotted line (blue) denotes $\beta
      F\sigma=25$ and $\phi=0.7431$, the dashed line (purple) $\beta
      F\sigma=500$ and $\phi=0.8025$, and the solid line (red) $\beta
      F\sigma=1000$ and $\phi=0.8042$.  The density evolves gradually
      over this range of $F$ and develops a noticeable shoulder by
      $\beta F\sigma=1000$, which anticipates the appearance of a
      secondary maximum.  On the right the dash-dotted line (blue)
      denotes $\beta F\sigma=1400$ and $\phi=0.8047$, the dashed line
      (purple) $\beta F\sigma=2000$ and $\phi=0.8053$, and the solid
      line (red) $\beta F\sigma=2500$ and $\phi=0.8058$.  The results
      show the development of a secondary maximum in the density, as
      the pressure is increased.  At infinite pressure, the center of
      the trapped disk [shaded in Fig.~\ref{fig:buckled}(a)] would be
      at $z=0.0840\,\sigma$.}
    \label{fig:density}
  \end{center}
\end{figure*}

If the apparent singularity in the force at $\phi_K$ is avoided by an
evolution to the buckled state, we should expect to see evidence for
this in $\rho_1(y)$, the probability density for finding a disk
at~$y$, defined in Eq.~\eqref{eq:rho1def}.  In Fig.~\ref{fig:density},
$\rho_1$ is plotted as a function of distance from one wall of the
channel containing the disk centers.  In all cases, the distribution
has its maximum at the wall, but a secondary maximum starts to appear
for $F>1000$, and is well developed for $F=2500$.  This secondary
maximum clearly corresponds to the trapping of the shaded disks in
Fig.~\ref{fig:buckled}(a): in the jammed state, the centers of these
disks are at a small distance (roughly $0.0840\,\sigma$ for
$h=0.95\,\sigma$) from the walls $y=\pm h/2$ that confine the centers
of the disks.

\section{Correlation Lengths}
\label{sec:corrlengths}

It is possible to obtain a great many correlation lengths from the
ratios of eigenvalues of the transfer integral equation.  In our work
we focus on three whose interpretation is particularly simple.

The longest correlation length is that associated with the growth of
zigzag order or, more precisely, the decay of the correlation between
the $y$-coordinates of well-separated disks $i$ and $i+s$,
\begin{equation}
  \langle y_i\, y_{i+s} \rangle \sim (-1)^s \exp(-s/\xi_{zz}),
\label{eq:zigzagdef}
\end{equation}
where
\begin{equation}
  \xi_{zz}=1/\ln(\lambda_1/|\lambda_2|)\,.
\end{equation}
The eigenfunction $u_1$ is of even parity (it is nodeless) and $u_2$
has odd parity, so that $y$ has a nonzero matrix element between $u_1$
and $u_2$; it may also be noted that the eigenvalues $\lambda_1$ and
$\lambda_2$ have opposite sign.  These two features are consistent
with the zigzag correlations described by Eq.~\eqref{eq:zigzagdef}.
The $F$-dependence of the length $\xi_{zz}$ is given in
\begin{figure}
  \begin{center}
    \includegraphics[width = 3.5in]{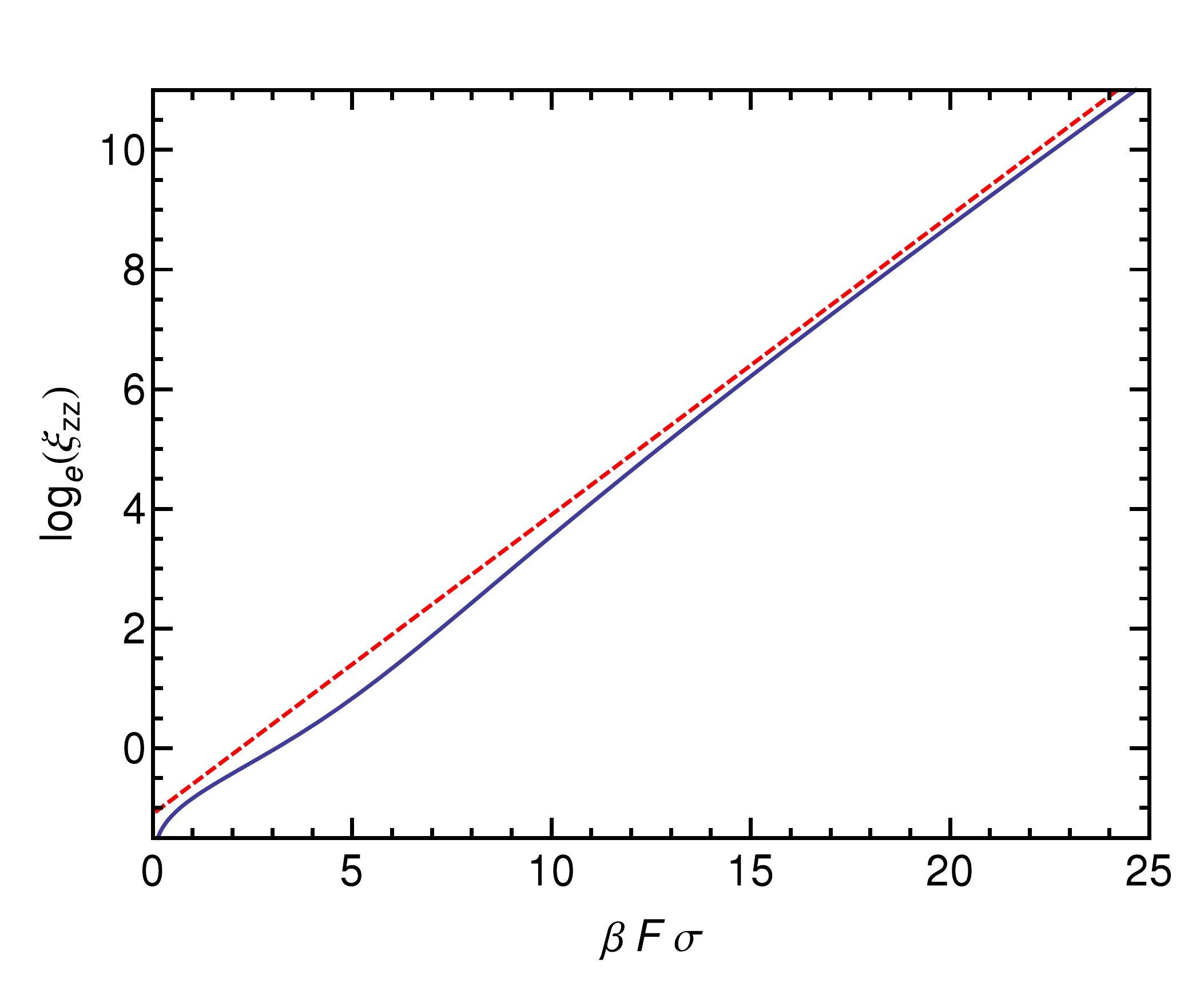}
    \caption{(Color online) The solid line denotes the correlation
      length $\xi_{zz}=1/\ln(\lambda_1/|\lambda_2|)$ derived from the
      second real eigenvalue $\lambda_2$.  It is a measure of the
      growth of zigzag order [see Eq.~\eqref{eq:zigzagdef}].  The
      dashed line is a straight line of gradient 0.5 [see
        Sec.~\ref{sec:lengthscalesexplained}].}
    \label{fig:zigzagxi}
  \end{center}
\end{figure}
Fig.~\ref{fig:zigzagxi} and is discussed in
Sec.~\ref{sec:lengthscalesexplained}.

The character of the eigenvalues $\lambda_n$ for $n>2$ depends on
$\beta F \sigma$.  As $\phi \to \phi_{\rm max}$ the next four
eigenvalues are two different complex conjugate pairs, as we shall
explain in Sec.~\ref{sec:highdensity}.  Outside this limit (and when
$\beta F\sigma$ is not small) the first few eigenvalues are real and
occur in parity-related doublets of opposite sign, just like
$\lambda_1$ and~$\lambda_2$.

\begin{figure}
  \begin{center}
    \includegraphics[width = 3.3in]{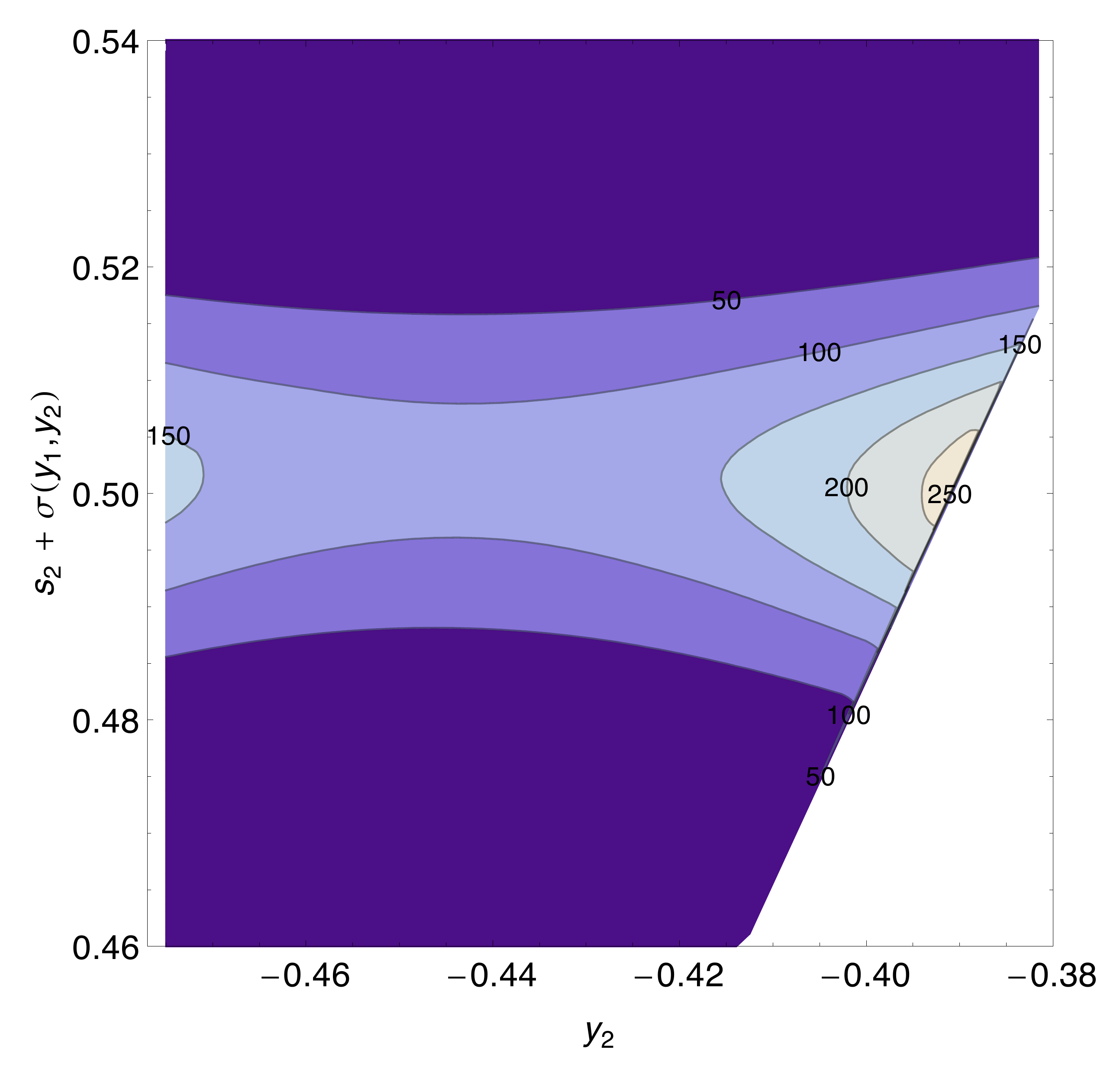}
    \caption{(Color online) Section through the nodeless eigenfunction
      $u_1(y_2,y_1,s_2)$ at $y_1=h/2$, for $h=0.95\,\sigma$ and $\beta
      F\sigma=1000$.  The eigenfunction is normalized so that
      $B[u_1,u_1]=1$ and has been scaled by $\exp(\beta
      F[s_2+\sigma_{1,2}]/2)$ for plotting.  Distance scales are in
      units of~$\sigma$. The white region in the lower right-hand
      corner corresponds to a disallowed region $s<0$, in which disks
      would overlap.}
    \label{fig:mode1}
  \end{center}
\end{figure}

\begin{figure}
  \begin{center}
    \includegraphics[width = 3.3in]{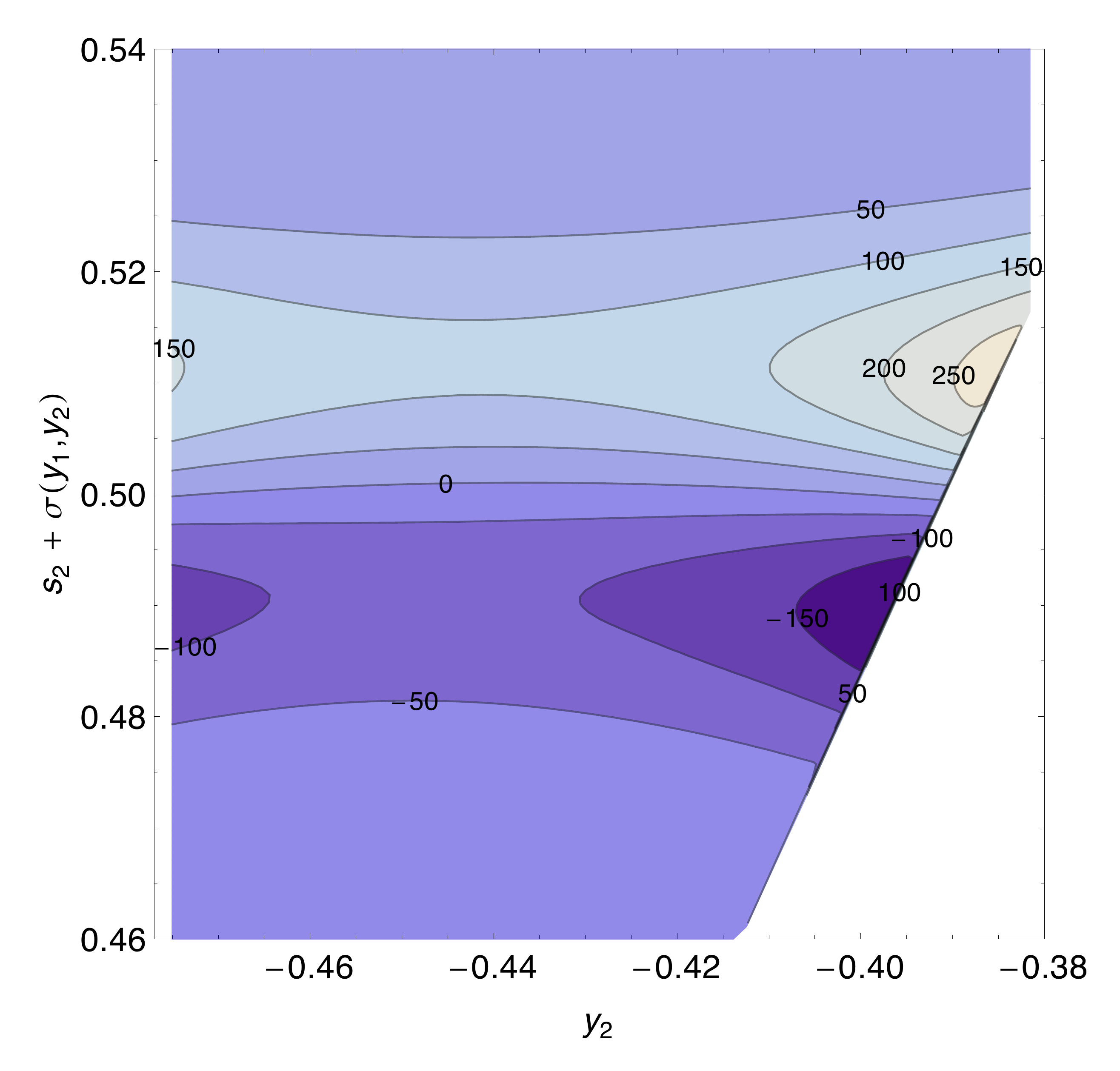}
    \caption{(Color online) Section through the second even-parity
      eigenfunction $u_3(y_2,y_1,s_2)$ at $y_1=h/2$, showing the nodal
      line at $s_2+\sigma_{1,2}\simeq0.50\,\sigma$.  The scaling of
      $u_3$ is the same as for $u_1$ in Fig.~\ref{fig:mode1}.}
    \label{fig:mode3}
  \end{center}
\end{figure}

A second length scale with a simple interpretation is $\xi_3= 1/\ln
(\lambda_1/ |\lambda_3|)$, calculated from the eigenvalues of the
first two even-parity eigenfunctions, $u_1$ and~$u_3$.  The functions
$u_1$ and $u_3$ are depicted in Figs.~\ref{fig:mode1} and
\ref{fig:mode3}
for the case $\beta F\sigma=1000$, for which the correlation length
$\xi_3\simeq26$.  The function $u_3$ has an approximately planar nodal
surface near $s_2+\sigma_{1,2}=0.50\,\sigma$, so it is clear that the
NN separation $x_2-x_1$ will have a large matrix element between $u_3$
and the nodeless function~$u_1$.  Accordingly, $\xi_3$ is the distance
over which the nearest-neighbor separations $x_{k+1}-x_k$ remain
correlated.

\begin{figure}
  \begin{center}
    \includegraphics[width = 3.5in]{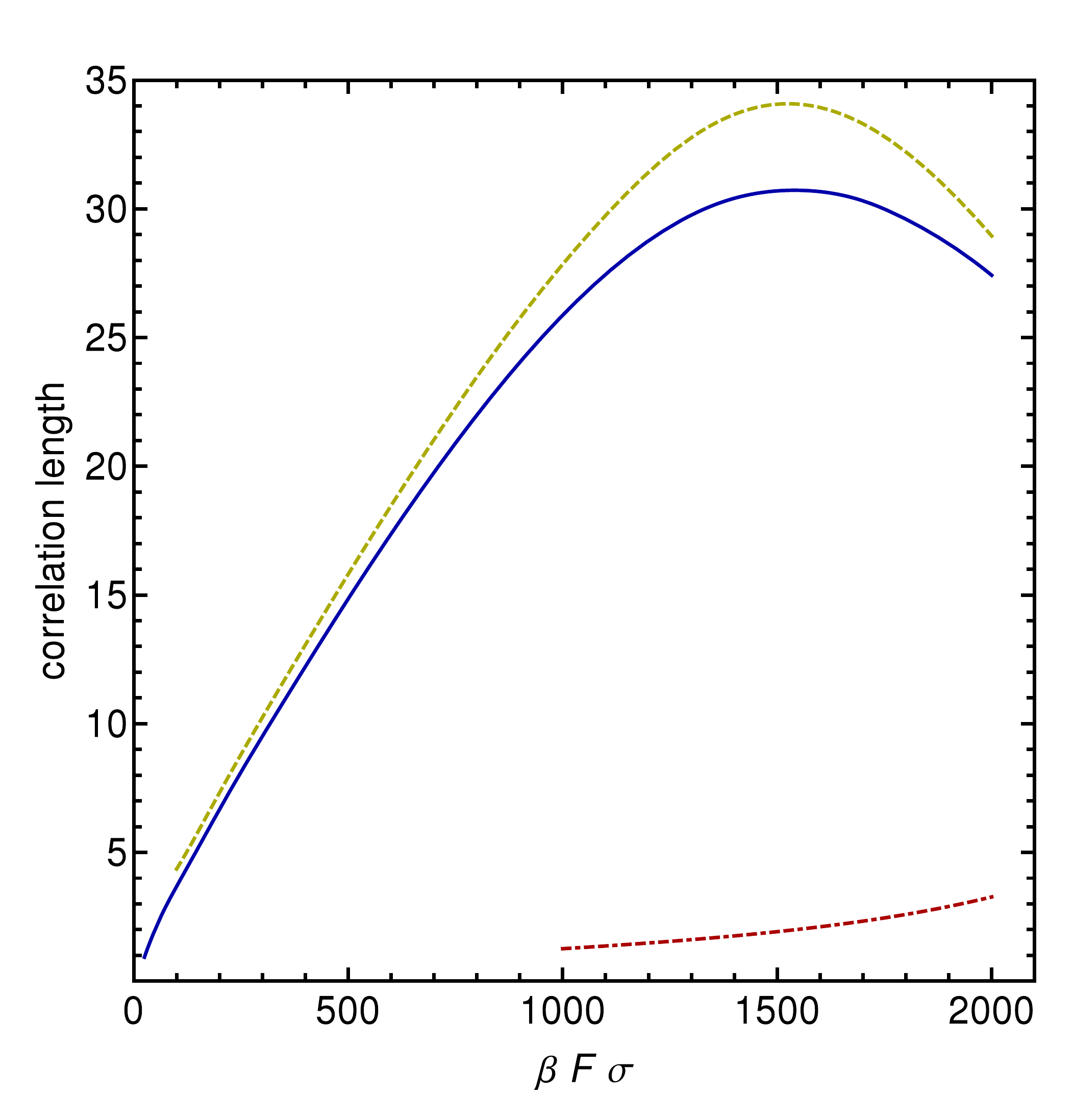}
    \caption{(Color online) Correlation lengths
      $\xi_3=1/\ln(\lambda_1/|\lambda_3|)$ (solid line) and
      $\xi_c=1/\ln(\lambda_1/|\lambda_c|)$ (dash-dotted line), derived
      from the third real eigenvalue $\lambda_3$ and the first complex
      eigenvalue~$\lambda_c$.  Here $\xi_3$, which has its greatest
      value for $\beta F\sigma\approx 1540$ (corresponding to
      $\phi\approx 0.8049$), is the length scale over which
      correlations between the separations of nearest-neighbor disks
      along the $x$-axis persist.  Further, $\xi_c$ describes the
      persistence of buckled-crystal order; this second length scale
      grows, but remains small in the range of $\beta F\sigma$ shown.
      Also shown is the approximation $\xi_3\simeq2\var{x_{\rm
          nnn}}/\var{x_{\rm nn}}$ (dashed line), discussed in
      Sec.~\ref{sec:lengthscalesexplained}.}
    \label{fig:corrlens}
  \end{center}
\end{figure}

It is shown in Fig.~\ref{fig:corrlens} that the correlation length
$\xi_3$ peaks at a value of around $30$ for the packing fraction $\phi
\approx 0.8049$.  Note that our $\xi_{zz}$ and $\xi_3$ are actually
\emph{dimensionless} quantities, as may be seen from
Eq.~\eqref{eq:zigzagdef}, for example.  To obtain the physical length
scales one has simply to multiply them by $L/N$.  Thus
\begin{equation}
  \tilde{\xi}_3\equiv\xi_3 L/N=\pi \sigma^2/(4 H_d \,\phi),
\label{units}
\end{equation}
where $\tilde{\xi}_3$ is the physical length.  When $H_d
=1.95\,\sigma$ and $\phi=0.8049$, $\tilde{\xi}_3 \approx
0.500\,\xi_3\sigma$.  The maximum value of $\xi_3$ shown in
Fig.~\ref{fig:corrlens} therefore corresponds to a physical length
scale $\tilde\xi_3\approx15\,\sigma$.

A correlation length that increases to a maximum and
then decreases is indicative of an avoided transition.  The value of
$\phi$ that corresponds to the peak in $\xi_3$ is a little less than
the value, $\phi_K=0.8054$, at which $(\beta F \sigma)^{-1}$ appears
to approach zero (see Fig.~\ref{fig:onebyF}).  Given that there is no
true phase transition at $\phi_K$, it is perhaps not surprising that
different measures for locating the underlying avoided transition
should disagree slightly.

The third length scale, $\xi_c$, that we have studied is a measure of
the extent of buckled-crystal order.  For the larger values of $\beta
F\sigma$, it can be determined from the ratio $\lambda_1/|\lambda_c|$,
where $|\lambda_c|$ is the magnitude of the largest complex conjugate
pair of eigenvalues (see Sec.~\ref{sec:highdensity}).
Figure~\ref{fig:corrlens} shows that $\xi_c$ grows with increasing
$\beta F\sigma$, whilst remaining small in the range of $\beta F
\sigma$ that we can study numerically.  Its expected behavior for
$\phi\to\phi_{\rm max}$ is discussed later, in
Sec.~\ref{sec:highdensity}.

\section{Probability Densities}
\label{sec:probdensities}

\begin{figure}
  \begin{center}
    \includegraphics[width = 3.5in]{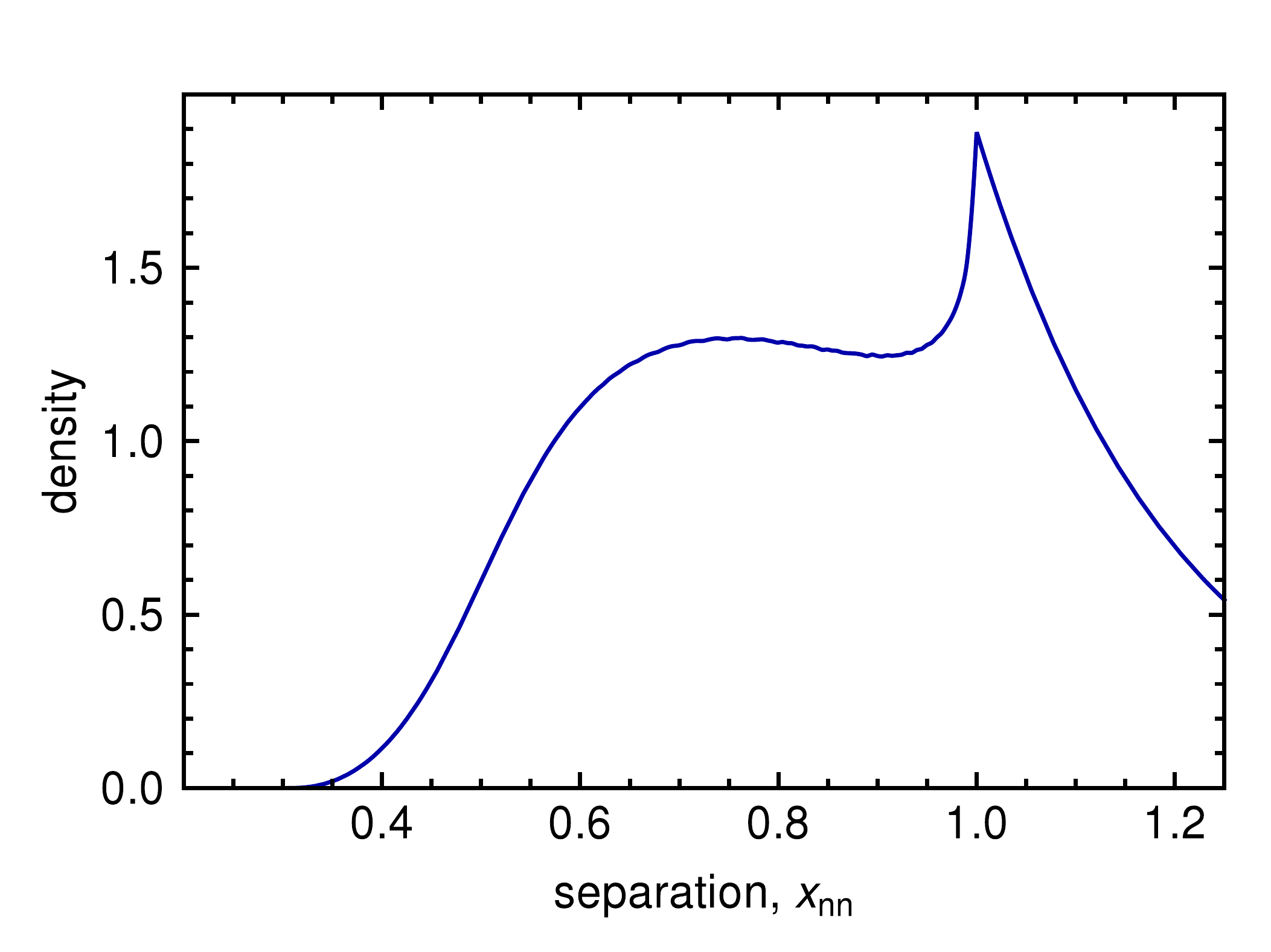}
    \caption{(Color online) Probability density for the centers of
      neighboring disks to be separated by distance $x_{\rm nn}$ along
      the channel, illustrated for $\beta F\sigma=5$ and
      $h=0.95\,\sigma$; $x_{\rm nn}$ is in units of~$\sigma$.  For
      small $F$, as shown here, the results are qualitatively similar
      to those found in \cite{GurinVarga} for a narrow channel with
      $h=\sigma\sqrt3/2$; the height of the sharp maximum at $x_{\rm
        nn}=\sigma$ decreases rapidly with increasing $F$.}
    \label{fig:denx}
  \end{center}
\end{figure}

\begin{figure}
  \begin{center}
    \includegraphics[width = 3.5in]{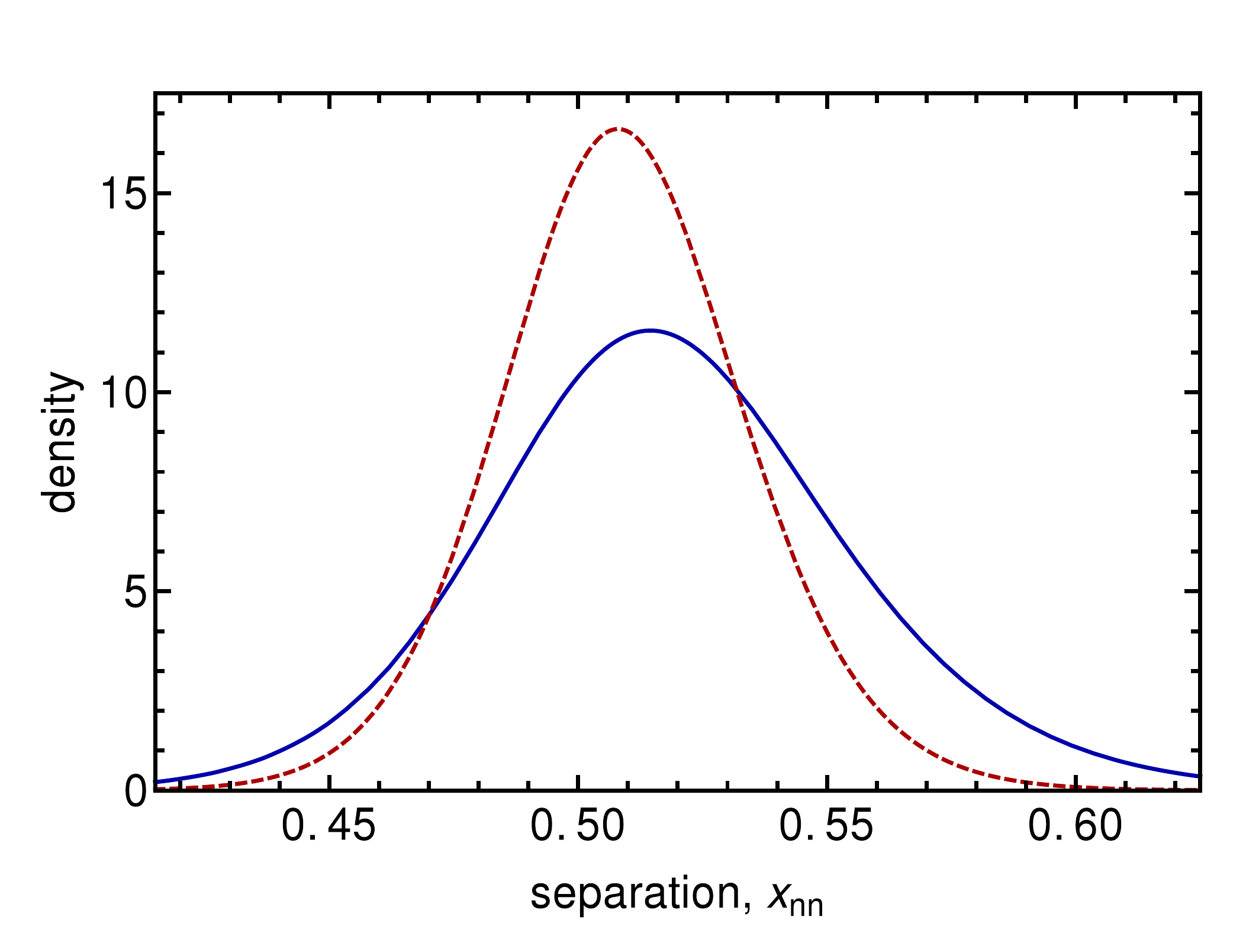}
    \caption{(Color online) Probability density for the centers of
      neighboring disks to be separated by distance $x_{\rm nn}$ along
      a channel with $h=0.95\,\sigma$, illustrated for $\beta
      F\sigma=50$ (solid line) and $100$ (dashed line); $x_{\rm nn}$
      is in units of~$\sigma$.  As $F$ increases, the distribution
      approaches the Gaussian form with a variance that decreases as
      $1/F$.}
    \label{fig:moredenx}
  \end{center}
\end{figure}

\begin{figure}
  \begin{center}
    \includegraphics[width = 3.5in]{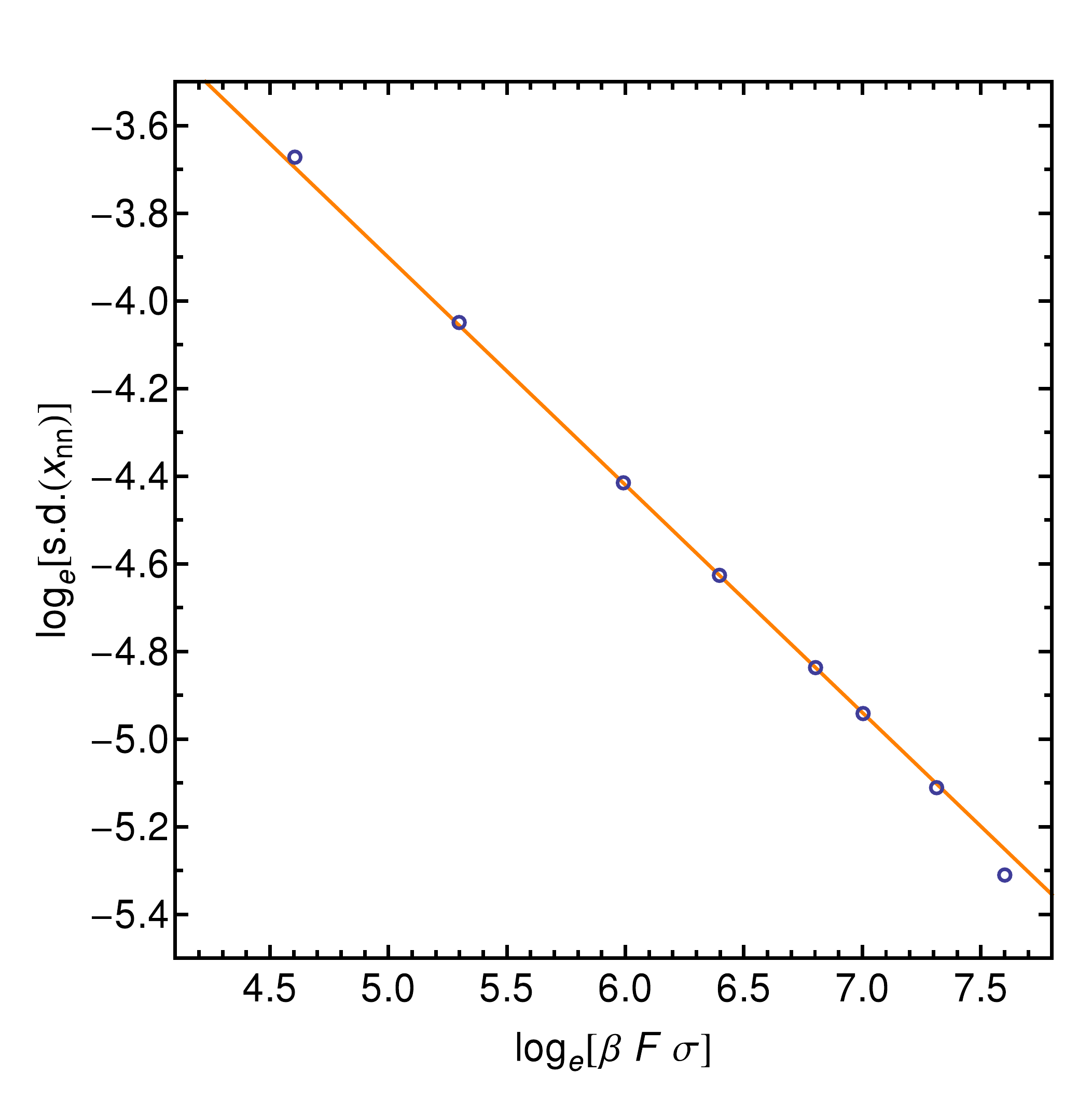}
    \caption{(Color online) Standard deviation of the nearest-neighbor
      separation plotted against $\beta F\sigma$; the scales are
      logarithmic.  The solid line has gradient $-0.52$, showing that
      the fluctuations in the nearest-neighbor distance decrease
      approximately as $1/\sqrt{F}$.}
    \label{fig:sd-xnn}
  \end{center}
\end{figure}

We have studied the probability density $\rho_{\rm nn}(x_{\rm nn})$
for the centers of neighboring disks to be separated by a distance
$x_{\rm nn}$ using
\begin{align}
  &\rho_{\rm nn}(x_{\rm nn}) \nonumber\\
  &\quad= \inti ds \inth dy_2 \inth dy_1
  \,\delta(x_{\rm nn}-s-\sigma_{2,1}) \rho(y_2,y_1,s).
\end{align}
This probability density changes in a striking manner as the packing
fraction or force $F$ is increased.  Figure~\ref{fig:denx} shows
$\rho_{\rm nn}$ for $\beta F \sigma =5$, which corresponds to a
packing fraction at which zigzag order is starting to develop.  For
small values of $\beta F \sigma$, the main feature is a cusp at
$x_{\rm nn} =\sigma$; but, as zigzag order grows, a broad hump
develops at a smaller value of $x_{\rm nn}$: both features may be seen
in Fig.~\ref{fig:denx}.  Results for larger values of $\beta F\sigma$
are shown in Fig.~\ref{fig:moredenx}.  When $\beta F \sigma = 50$ the
cusp at $x_{\rm nn}=\sigma$ has nearly disappeared and would be
undetectable on the scale used in Fig.~\ref{fig:moredenx}.  As $\beta
F \sigma$ increases further, the position of the new peak moves
towards the value $x_{\rm nn}\simeq\sigma/2$, as would be expected
from Fig.~\ref{fig:buckled}(b).  For $\beta F \sigma =100$, the peak
is nearly Gaussian in shape and we have directly determined its width
by obtaining the variance $\langle (x_{\rm nn}-\langle x_{\rm
  nn}\rangle)^2\rangle$.  The standard deviation of the
nearest-neighbor separation is plotted in Fig.~\ref{fig:sd-xnn}: it
varies as $1/\sqrt{F}$ as $\beta F \sigma$ increases, which is a much
slower decrease than would have been naively expected and reflects the
ease of sliding the top row with respect to the bottom row in
Fig.~\ref{fig:buckled}(b).

The separation of next-nearest neighbors has much smaller
fluctuations.  The variance of $x_{\rm nnn}\equiv(x_{k+2}-x_k)$ is
given by
\begin{align}
  \var{}&x_{\rm nnn} \nonumber\\
  &= \var([x_{k+2}-x_{k+1}]+[x_{k+1}-x_k])\nonumber\\
  &= 2\var x_{\rm nn}+2\cov(x_{k+2}-x_{k+1},x_{k+1}-x_k)\,.
\label{eq:covar}
\end{align}
The covariance term (which is negative) can be obtained from~$u_1$.
By following a procedure similar to that described in
Sec.~\ref{sec:distribution}, it can be shown that
\begin{align}
  \langle (x_{k+2}-x_{k+1})(x_{k+1}-x_k)\rangle
  &= \frac{K[e_1,e_1]}{K[u_1,u_1]} \nonumber\\
  &= \frac{K[e_1,e_1]}{\lambda_1B[u_1,u_1]}\,,
\label{eq:covexpr}
\end{align}
where $e_1(y_2,y_1,s_2)=(s_2+\sigma_{2,1})\, u_1(y_2,y_1,s_2)$.  The
calculation of $\var x_{\rm nnn}$ from \eqref{eq:covar} and
\eqref{eq:covexpr} thus requires only the distribution of $x_{\rm nn}$
(obtained from $u_1$) and a single application of the transfer matrix
to~$e_1$.

\begin{figure}
  \begin{center}
    \includegraphics[width = 3.5in]{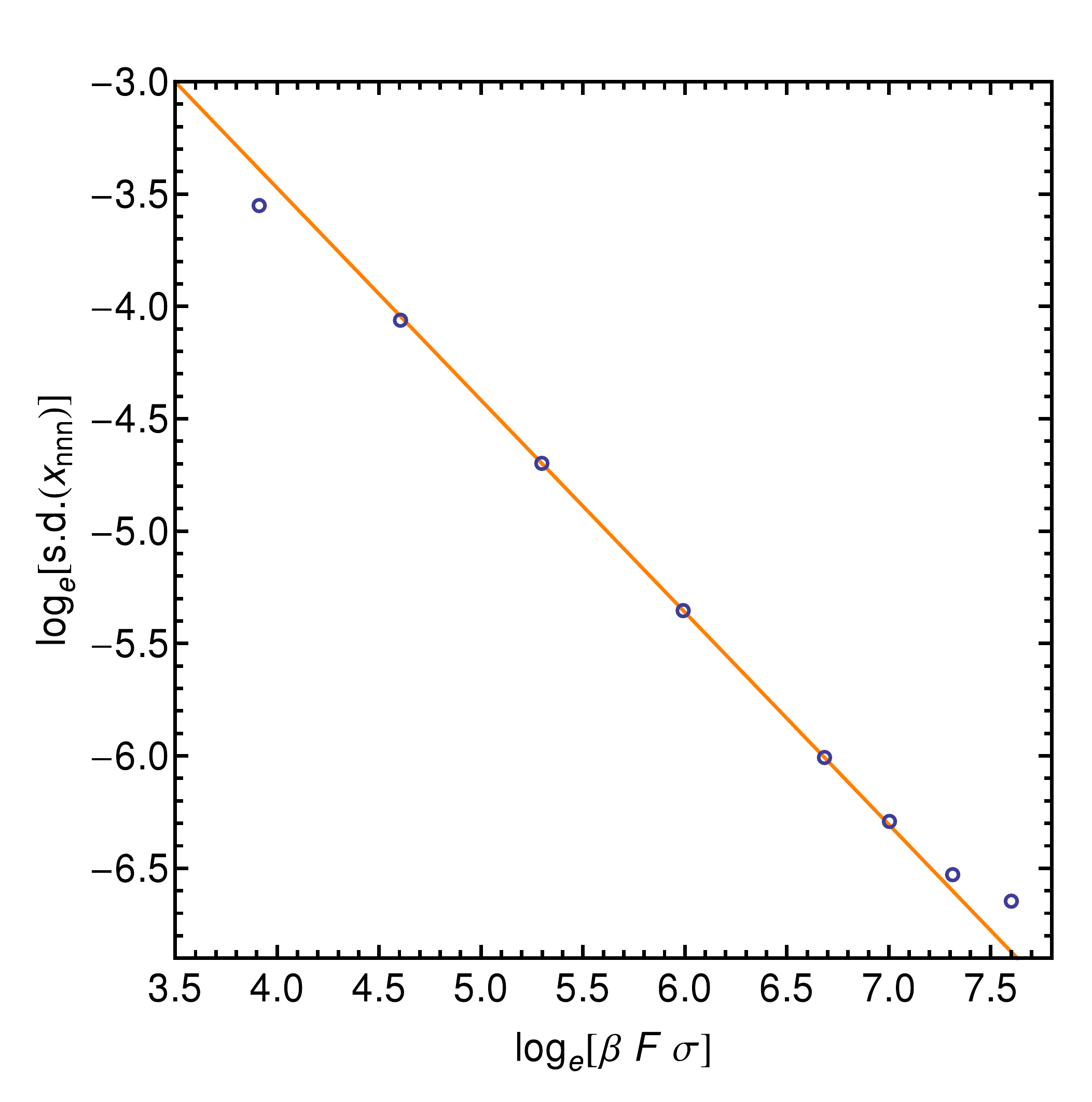}
    \caption{(Color online) Standard deviation of the separation of
      next-nearest-neighbor disks plotted against $\beta F\sigma$; the
      scales are logarithmic.  The solid line has gradient $-0.94$,
      showing that the fluctuations in the next-nearest-neighbor
      distance decrease approximately as $1/F$}
    \label{fig:sd-xnnn}
  \end{center}
\end{figure}

In Fig.~\ref{fig:sd-xnnn} the standard deviation of $x_{\rm nnn}$ is
plotted for a range of $F$ up to $\beta F\sigma=2000$.  It is close to
the expected $1/F$ form, but with some flattening off at the largest
values of~$F$.  Eventually, the standard deviation must tend to a
constant, independent of $F$, because, as the state of highest density
is approached, the NNN separation can have either of two distinct
values (see Fig.~\ref{fig:buckled}(a)), with fluctuations of order
$1/F$ around them.

\section{Understanding the correlation lengths}
\label{sec:lengthscalesexplained}

There are two important length scales in our problem: the length scale
associated with the growth of zigzag or bond-orientational order,
$\xi_{zz}$ (see Eq.~\eqref{eq:zigzagdef}), and the length scale
$\xi_3$, which describes how the correlations between the
nearest-neighbor separations of pairs of disks at numbers $k$ and  $k+s$
decay with $s$,
\begin{equation}
  \langle (x_{k+1}-x_k)(x_{k+s+1}-x_{k+s})\rangle_c
  \sim (-1)^se^{-s/\xi_3}\,,
\label{eq:xi-def}
\end{equation}
where $c$ denotes the cumulant of the correlation.
$\xi_3$ is a measure of the distance along the channel over which the
system with $\phi\simeq\phi_K$ comes to resemble
Fig.~\ref{fig:buckled}(b).  We will show later that $\xi_3$ can also
be regarded as the size of a localized thermal fluctuation in which
disks of one row are displaced relative to the other row.  We have
previously obtained $\xi_{zz}$ and $\xi_3$ from the eigenvalues of the
transfer matrix, but in this section we give simple arguments to
explain their dependence on~$F$.

In Ref.~\cite{Godfrey} we were able to give an account of the
dependence of $\xi_{zz}$ for the NN model by relating it to the
typical distance between defects.  For the NNN model, there are many
types of defects~\cite{AshwinBowles}, but we believe the following
argument captures the physics of what is going on.

For moderate values of $F$, typical configurations of the system
resemble Fig.~\ref{fig:buckled}(b).  A pair of defects in this zig-zag
pattern can be created by removing one disk from the pattern and
reinserting it elsewhere, between two disks that lie on the same side
of the channel.  If the two rows of disks maintain their relative
position between the two defects, the increase in length of the system
will be $\sigma$ \cite{NNDelta}, giving an additional length per
defect $\Delta_a\simeq\sigma/2$; the work done to create it will be
$F\sigma/2$.  We therefore expect the density of defects to vary as
$\exp[-\beta F\sigma/2]$ and the correlation length $\xi_{zz}$ to vary
as $\exp[\beta F\sigma/2]$, which is the typical spacing of the
defects.  The dashed line in Fig.~\ref{fig:zigzagxi} shows that this
simple argument gives a reasonably good account of the numerical data
for $\beta F \sigma < 25$, despite its neglect of any relaxation of
the system near the defects.  We also note that agreement cannot be
maintained indefinitely, as our argument ignores the eventual
appearance of buckled crystalline order.

The length scale $\xi_3$ has no obvious connection with defects, but
instead can be understood as an effect due to the accumulation of the
random variations in the spacing of NNN disks.  To estimate $\xi_3$,
we assume that the coupling of the two rows of disks is weak and that
the gaps between next-nearest neighbors are approximately independent
random variables, like the gaps between neighbors in a one-dimensional
gas of hard rods.  With the notation $x_{i,j}\equiv x_i-x_j$ we can
write
\begin{equation}
  \sum_{k=1}^m (-1)^k x_{k+1,k-1} = (-1)^m x_{m+1,m} - x_{1,0}\,.
\end{equation}
The left-hand side of this identity is an alternating sum of $m$
next-nearest-neighbor separations: to the extent that these
separations are independent random variables, the variance of the
left-hand side will grow as $m\var x_{\rm nnn}$ as $m$ increases.  At
the same time, the variance of the right-hand side will increase
towards $2\var x_{\rm nn}$ as the correlation between $x_{m+1,m}$ and
$x_{1,0}$ decreases.  We therefore expect the correlation to be small
when $m\var x_{\rm nnn} \simeq 2\var x_{\rm nn}$.  By setting
$m=\xi_3$ we obtain the estimate
\begin{equation}
  \xi_3\simeq 2\,\frac{\var x_{\rm nn}}{\var x_{\rm nnn}}.
\label{eq:xi3estimate}
\end{equation}
Fig.~\ref{fig:corrlens} shows how well this simple formula works.  It
accounts for both the increase of $\xi_3$ to its maximum as a function
of $F$ as well as its decrease.

The correlation length $\xi_3$ can also be understood as the typical
number of disks that participate in a thermal fluctuation in which a
portion of one row moves relative to the other row.  The configuration
shown in Fig.~\ref{fig:buckled}(c) could be regarded as the most
extreme fluctuation of this kind: it involves the correlated motion of
many disks, and so should be expected to be long-lived on the scale of
the collision time.  In a localized fluctuation in which $m$ disks of
one row are displaced relative to the other, the distances between the
disks in one row (averaged over a time that is short compared with the
duration of the fluctuation) will change by small amounts of
order~$\pm\Delta$, with $\Delta \sim (\var x_{\rm nn})^{1/2}/m$.  The
total strain energy, which is on the order of the thermal energy
$k_BT$, will be approximately
\begin{equation}
  m \times\kappa\, \Delta^2 \sim k_BT\,,
\label{eq:strainenergy}
\end{equation}
where the effective spring constant $\kappa$ for the force
conjugate to the next-nearest neighbor separation can be estimated
from the fluctuation--response relation
\begin{equation}
  \kappa^{-1} = \beta\var x_{\rm nnn}\,.
\label{eq:kappa}
\end{equation}
With our estimate for $\Delta$, Eqs.~\eqref{eq:strainenergy} and
\eqref{eq:kappa} lead to $m\sim\xi_3$, where $\xi_3$ is given by
Eq.~\eqref{eq:xi3estimate}.

\section{Behavior as \texorpdfstring{$\phi\to\phi_{\rm max}$}{the density tends to a maximum}}
\label{sec:highdensity}

We have been unable to obtain numerical solutions of the integral
equation \eqref{eq:integraleq} for very large values of the force,
$\beta F \sigma>2500$.  To help fill this gap left by our numerical
work, in this section we present an analytical method of solution
which is expected to become exact in the limit $\beta F \sigma \to
\infty$.  We relate the period-six correlations of the buckled crystal
to the asymptotic form of the eigenvalues and we calculate the
correlation length for crystalline order.  We also indicate what
features of the eigenfunctions will lead to the high-density form of
the equation of state given in Eq.~\eqref{eq:SW}.

For $F\to\infty$, the system of disks adopts the high-density buckled
crystal configuration shown in Fig.~\ref{fig:buckled}(a), but for any
finite $F$ the crystalline order will be disrupted by the presence of
mobile defects.  The concentration of these defects will vary as
$\exp(-\beta F\Delta_d)$, where $\Delta_d$ is the additional length
needed to accommodate a defect in the crystal.  We therefore expect
the mean spacing of defects to increase with $F$ according to
\begin{equation}
  \xi_c \sim \exp(\beta F\Delta_d)\,.
\label{eq:xid}
\end{equation}
Our notation for this length scale reflects the fact that $\xi_c$ will
also be the distance over which crystalline order persists in the
presence of defects.

\begin{figure}
  \begin{center}
    \includegraphics[width = 3.5in]{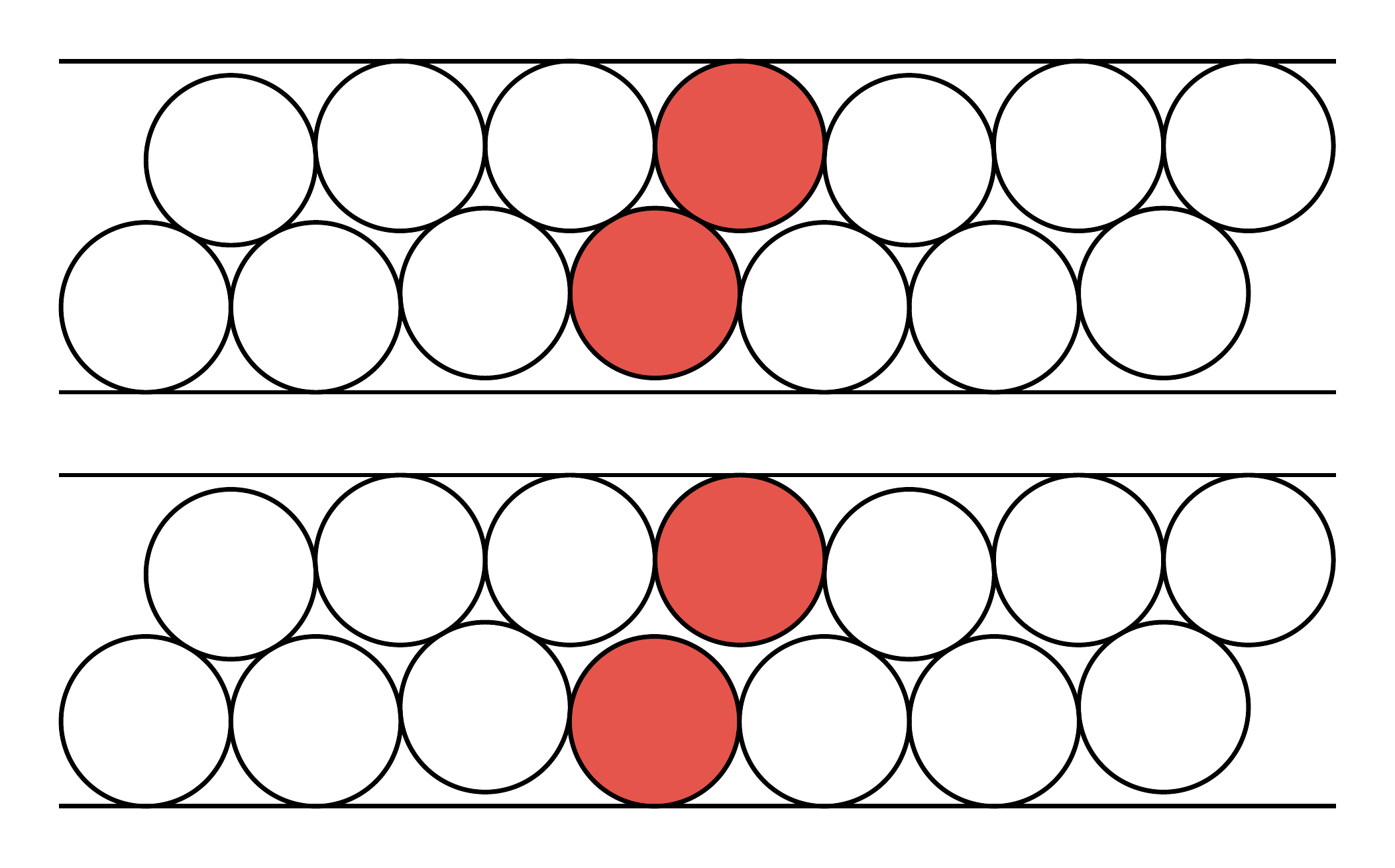}
    \caption{(Color online) Defects in the high-density
      buckled-crystal state illustrated in Fig.~\ref{fig:buckled}(a).
      Two distinct types of defect are shown: Each can be created by
      inserting disks into the buckled crystal at the positions shown
      (shaded) or at positions related to these by symmetry.}
    \label{fig:defects}
  \end{center}
\end{figure}

The shortest (and hence most abundant) defects are of two kinds,
illustrated in Fig.~\ref{fig:defects}.
Each kind of defect can be created by inserting a pair of disks into
the buckled crystal.  If six disks, comprising one complete unit cell
of length $a$ in the buckled crystal, are removed from a region of
perfect crystal, these disks can be reinserted elsewhere in the
crystal to create three defects.  The net increase in length is
\begin{equation}
  3\Delta_d = 3\sigma - a\,,
\end{equation}
where $a=\sigma+2\sigma_\epsilon$ and $\sigma_\epsilon=[\sigma^2 - (h -
  \sigma\sqrt3/2)^2]^{1/2}$, so that
\begin{equation}
  \Delta_d = \tfrac23(\sigma-\sigma_\epsilon)\,.
\label{eq:deltad}
\end{equation}

We believe it is worth understanding how the correlation length
$\xi_c$ given by \eqref{eq:xid} and \eqref{eq:deltad} can arise from
the integral equation \eqref{eq:integraleq} and also how the
periodicity of the buckled crystal can be explained.  These two
matters are very closely connected: the periodicity must appear in the
correlation functions, and the dominant contributions to the
large-distance correlation functions can be constructed from the
eigenfunctions whose eigenvalues are closest in magnitude to the
largest eigenvalue~$\lambda_1$.  It follows that there should, for
sufficiently large $F$, be \emph{complex} eigenvalues of
\eqref{eq:integraleq} whose magnitudes are very similar to~$\lambda_1$
and whose phases can be related to the periodicity of the buckled
crystal.  In what follows we show that complex eigenvalues with these
properties can be inferred from the form of the integral
equation~\eqref{eq:integraleq}.

For sufficiently large $F$, the solutions of \eqref{eq:integraleq}
will be sharply peaked near $y=\pm h/2$ and $\pm(\sqrt3\,\sigma-h)/2$,
with peak widths proportional to $(\beta F)^{-1}$.  As seen in
Fig.~\ref{fig:density}, this transverse localization is starting to
appear above $\beta F\sigma=1000$ and it is well developed when $\beta
F\sigma=2500$.  But in addition to this localization in $y_1$ and
$y_2$, the eigenfunctions will also eventually show a similar degree
of localization in~$s_2$.  In the buckled crystal (see
Fig.~\ref{fig:buckled}(a)), adjacent disks are either touching, with
$s_2=0$, or non-touching, with
$s_2=\sigma_\epsilon-\frac12\sigma-\sigma_h$, where
$\sigma_h=\sqrt{\sigma^2-h^2}$.  In addition to these two separations,
the shaded disks in the second of the configurations shown in
Fig.~\ref{fig:defects} are separated by a third distance,
$s_2=\frac32\sigma-\sigma_\epsilon-\sigma_h$.  This local information
regarding pairs of disks can be used to contruct an approximate
solution of the integral equation.

From the preceding discussion, we expect a solution $u(y_2,y_1,s_2)$
of \eqref{eq:integraleq} to have eight narrow peaks with widths
${\sim}(\beta F)^{-1}$.  If $u$ has even parity, there will be only four
distinct peak values of $u$, which we denote by
\begin{equation}
  \begin{aligned}
    v_1 &= u\bigl(-\tfrac12h,\tfrac12h,
    \tfrac32\sigma-\sigma_\epsilon-\sigma_h\bigr) \\
    v_2 &=  u\bigl(-\tfrac12h,\tfrac12h,
    \sigma_\epsilon-\tfrac12\sigma-\sigma_h\bigr) \\
    v_3 &= u\bigl(-\tfrac12h,\tfrac12h-\epsilon,0\bigr) \\
    v_4 &= u\bigl(-\tfrac12h+\epsilon,\tfrac12h,0\bigr)\,,
  \end{aligned}
\end{equation}
where $\epsilon$ is given by~\eqref{eq:epsdef}.  Approximate equations
satisfied by the quantities $v_1$ to $v_4$ can be read off from
Eq.~\eqref{eq:integraleq}; for example,
\begin{equation}
  \lambda\,v_2 \simeq e^{-\beta F(\sigma_\epsilon-\frac12\sigma)}\,(q_1v_1+q_2v_3)\,.
\label{eq:approxv2}
\end{equation}
In Eq.~\eqref{eq:approxv2}, the factors $q_1$ and $q_2$ arise from the
integrations over $y_0$ and $s_1$ in Eq.~\eqref{eq:integraleq}, so
they will be of order $(\beta F)^{-2}$ for sufficiently large~$F$.
The exponential factor, which corresponds to $\exp(-\beta
F[s_2+\sigma_{2,1}])$ in \eqref{eq:integraleq}, can be transferred to
the left-hand side of~\eqref{eq:approxv2}.  By proceeding in this way
for the remaining quantities $v_i$, we obtain a matrix eigenvalue
equation
\begin{equation}
  \lambda \mathsf{B}\mathbf{v} = \mathsf{K}\mathbf{v}\,,
\label{eq:matrixev}
\end{equation}
where $\mathbf{v}$ is a vector with components $v_i$ and the matrices
$\mathsf{K}$ and $\mathsf{B}$ are given by
\begin{equation}
  \mathsf{K} =
  \begin{pmatrix}
    0 & q_1 & 0 & 0 \\
    q_1 & 0 & q_2 & 0 \\
    0 & q_2 & q_3 & 0 \\
    0 & 0 & 0 & q_4
  \end{pmatrix}
\end{equation}
and
\begin{equation}
  \mathsf{B} =\frac1{\sqrt k}
  \begin{pmatrix}
    k_\epsilon/k & 0 & \bw0 & \bw0 \\
    0 & k/k_\epsilon & 0 & 0 \\
    0 & 0 & 0 & 1 \\
    0 & 0 & 1 & 0
  \end{pmatrix}
  ,
\end{equation}
where $k=\exp(-\beta F\sigma)$ and $k_\epsilon=\exp(-\beta
F\sigma_\epsilon)$.

The symmetry of the matrix $\mathsf{K}$ is not obvious from the
intuitive derivation given here.  To refine the argument, it can be
supposed that the eigenfunction $u$ is approximated by a linear
combination $u=\sum_i v_ib_i$ of localized functions
$b_i(y_2,y_1,s_2)$.  Projection of Eq.~\eqref{eq:integraleq} onto the
basis functions then leads to an equation of the form
\eqref{eq:matrixev}, in which the symmetry of the matrices
$\mathsf{B}$ and $\mathsf{K}$ follows from the symmetry of the
functionals $B[b_i,b_j]$ and $K[b_i,b_j]$ defined in
Eqs.~\eqref{eq:Bfunctional} and~\eqref{eq:Kfunctional}.  It might
therefore be an interesting task to construct a localized basis
adapted to the special case $\phi\to\phi_{\rm max}$, but we do not
attempt this here.  Instead we proceed to the qualitative results that
can be derived from the matrix eigenvalue
equation~\eqref{eq:matrixev}.

The equation for $\lambda$, obtained from $\det(\mathsf{K} -
\lambda\mathsf{B}) = 0$, is
\begin{equation}
  \lambda^4
  -(q_1^2+q_3q_4) k\lambda^2
  - q_2^2q_4 \sqrt k\,k_\epsilon \lambda
  + q_1^2q_3q_4k^2
  = 0\,.
\label{eq:secular}
\end{equation}
To determine the form of the solutions of this equation, it is helpful
first to introduce a scaled, dimensionless variable $\mu =
\lambda/(q_2^2\,q_4\,\sqrt k\,k_\epsilon)^{1/3}$.  Written in terms of
$\mu$, \eqref{eq:secular} becomes
\begin{equation}
  \mu^4
  - p_1r\mu^2
  - \mu
  + p_2 r^2
  = 0\,,
\label{eq:secularmu}
\end{equation}
where $r=(k/k_\epsilon)^{2/3}=\exp(-\beta F\Delta_d)$ is small and $p_1$ and
$p_2$ are expected to be independent of $F$ for large~$F$.

In the limit $r\to0$ (or $\beta F \Delta_d\to\infty$), the solutions
of \eqref{eq:secularmu} are $\mu=0$, 1, $\omega$, and~$\omega^2$, where
$\omega$ is a complex cube root of unity.  For small $r$, the
solutions can be expanded to first order in~$r$, giving
\begin{equation}
  \mu = 1+\tfrac13\,p_1r, \;
  \omega + \tfrac13\,p_1\omega^2\,r, \;
  \omega^2 + \tfrac13\,p_1\omega\,r
\label{eq:musoln}
\end{equation}
for the three roots with modulus close to unity; the fourth root is of
order~$r^2$.

So far we have considered only the even-parity solutions of
\eqref{eq:integraleq}.  The same method can be applied to the
odd-parity solutions, which leads to a further three roots $\mu$ with
modulus close to unity.  These roots are given by the same expressions
as in \eqref{eq:musoln}, but with the opposite signs.  This leads to a
two-fold degeneracy in $|\mu|$ for the real solutions and a four-fold
degeneracy for the complex solutions.  In our approximate treatment,
the cause of the degeneracy is our selective use of configurations in
which neighboring disks lie at (or near) opposite sides of the
channel, so that perfect zigzag order is assumed.  This assumption
about the configurations is reasonable, given our knowledge that
$\xi_{zz}\gg\xi_c$ for large $F$, but it does prevent us from
estimating the correlation length $\xi_{zz}$ from the ratio of the
first two real eigenvalues.

The result that emerges from our analysis is that, for sufficiently
large $F$, the six eigenvalues of largest magnitude are approximately
$(-\omega)^n\,\lambda_1$, with $n=0$ to~5.  An observable such as $y$
will have nonzero matrix elements between $u_1$ and the complex
conjugate pair of eigenfunctions with $\mu\simeq-\omega$
and~$-\omega^2$, which are the primitive sixth roots of unity.  The
correlation function $\langle y_i\,y_{i+s}\rangle$ will therefore
contain a contribution that oscillates with period six in~$s$.

To calculate the correlation length $\xi_c$ we must go beyond the
zeroth order in $r$ and use the expressions given in
Eq.~\eqref{eq:musoln}.  We find
\begin{align}
  \xi_c &\simeq 1/\ln([1+\tfrac13\,p_1r]/|\omega +
  \tfrac13\,p_1\omega^2\,r|) \nonumber\\
  &\simeq \frac2{p_1}\exp(\beta F\Delta_d)\,,
\end{align}
which is of the form anticipated in~\eqref{eq:xid}, though we have not
determined the quantity $p_1$, which will be a function only of the
ratio~$h/\sigma$.

Finally, we use the large-$F$ approximation
\begin{equation}
  \lambda_1 \simeq \bigl(q_2^2\,q_4\,\sqrt k\,k_\epsilon\bigr)^{1/3}
\end{equation}
to derive the equation of state for $\phi\to\phi_{\rm max}$.  The free
energy per disk is given by
\begin{align}
  \beta\Phi/N
  &= -\ln\lambda_1 \nonumber\\
  &= -\tfrac13\ln\bigl(q_2^2\,q_4\,\sqrt k\,k_\epsilon\bigr) \nonumber\\
  &= \tfrac16\beta F(\sigma + 2\sigma_\epsilon) + 2\ln(\beta F) + \mathrm{O}(1)\,,
\label{eq:Phiasym}
\end{align}
in which the $F$-dependence of $q_2$ and~$q_4$ has been used to obtain
the coefficient of the logarithmic term; the contribution written as
$\mathrm{O}(1)$ contains only terms that do not increase with~$F$.
Differentiation of $\Phi$ with respect to $F$
(cf.\ Eq.~\eqref{eq:LPhi}) gives the equation of state
\begin{equation}
  \beta F \simeq \frac{2N}{L-\frac16Na}
  = \frac{d_{\rm ef{}f}N}{L(1-\phi/\phi_{\rm max})}
\end{equation}
expected on the general grounds discussed in Ref.~\cite{SW}.  The
factor $d_{\rm ef{}f}=2$ in the expression for the force is the same
coefficient 2 that multiplies $\ln(\beta F)$ in
Eq.~\eqref{eq:Phiasym}.  In the context of our derivation, the source
of this factor is the two-fold integration in
Eq.~\eqref{eq:integraleq}, where the eigenfunction is localized within
regions whose widths are proportional to~$(\beta F)^{-1}$.

\section{Discussion}
\label{sec:conclusion}

We have solved the transfer integral equation appropriate to disks
moving in a channel of a width which permits NNN contacts and we have
determined its thermodynamic properties and some of its correlation
lengths numerically exactly.  The most striking feature of our results
are those connected with the features at $\phi_d$ and at~$\phi_K$.

The packing fraction $\phi_d$ marks the point at which
bond-orientational order (i.e.\ zigzag order) sets in, and will be
where the dynamics of the system will start to become activated.  We
believe that if the channel width were increased this feature would
evolve into the fluid--hexatic phase transition of hard
disks~\cite{KB}.  The new feature, which is absent for narrower
channels with only NN contacts, is that at $\phi_K$.  We believe it
would evolve for wider channels into the hexatic--crystal phase
transition~\cite{KB}.  It has a clear connection with the development
of structural order.

Despite being of structural origin, the behavior near $\phi_K$ mimics
that normally associated with the ideal glass transition.  At the
ideal glass transition, a divergence of the point-to-set length scale
is expected~\cite{Cammarota,Cavagna} and this will be accompanied by a
divergence of the penetration length for amorphous
order~\cite{Biroli}.  A review of some of the length scales which are
perhaps of relevance to amorphous systems can be found in
Ref.~\cite{LevineKurchan}.  This penetration length is the distance
over which the effect of a boundary condition (imposed by freezing the
positions of a subset of spheres) extends into the liquid; for the
case of three dimensions it has been determined from molecular
dynamics simulations by Gradenigo et al.~\cite{Cavagnaetal} and by
Berthier and Kob~\cite{KobBerthier}.  For our system of disks in a
channel, freezing the positions of the disks in the region $x<0$
supplies a boundary condition for the motion of the mobile disks in
the region~$x>0\,$: in particular, it fixes the nearest-neighbor
separation near $x=0$.  Following the discussion in
Secs.~\ref{sec:corrlengths} and~\ref{sec:lengthscalesexplained}, the
time-averaged nearest-neighbor separations $\langle
x_{k+1}-x_k\rangle$ will approach the bulk value $L/N$ as
$(-1)^k\exp(-k/\xi_3)$ for large $k$, so that we can identify the
penetration length with~$\xi_3$.

At the ideal glass transition it is expected that the relaxation time
$\tau_{\alpha}$ should diverge~\cite{Zamponi}.  It is argued in
Refs.~\cite{Godfrey,Josh} that $\tau_\alpha$ is the time at which a
disk can escape its cage by crossing from one side of the channel to
the other.  Within transition-state theory, whose accuracy for a
system of hard disks has been tested in Ref.~\cite{Barnett},
$\tau_\alpha$ is given by
\begin{equation}
  \tau_\alpha \sim \exp(\beta F \Delta_b),
\end{equation}
where the length $\Delta_b$ is $\mathrm{O}(\sigma)$, as discussed in
Ref.~\cite{Godfrey} for the case of nearest-neighbor interactions.  On
using Eq.~\eqref{eq:fittoSW} for $F$, the relaxation time will
\emph{appear} to diverge in a Vogel--Fulcher manner of
Eq.~\eqref{eq:VFform} for a range of $\phi$ less than~$\phi_K$.  Note
that at $\phi_K$ it will be a very challenging problem to study the
properties of the equilibrated system by molecular dynamics: the
densities are so close to $\phi_{\rm max}$ that the time scales in the
system will be extremely long.  Any real divergence of $\tau_\alpha$
will, of course, be avoided in the narrow channel systems, and we
suspect that it might also be avoided for hard spheres in three
dimensions~\cite{Moore,YeoMoore}.

Above the ideal glass transition discussed in the infinite dimensional
limit, i.e.\ when $\phi> \phi_K$, particles are expected to be caged
near their initial positions forever, whereas for $\phi < \phi_K$, the
particles can escape to infinity.  In our narrow channel system, the
particles are in one sense caged for all time as they cannot pass each
other; nevertheless, we have seen in Sec.~\ref{sec:probdensities} that
there is a qualitative difference in the nature of the confinement
above and below $\phi_K$.  Below $\phi_K$, the nearest-neighbor
separations in the $x$-direction fluctuate as $1/\sqrt{F}$, whereas
above $\phi_K$ the fluctuations are smaller and will decrease as $1/F$
as $\phi\to\phi_{\rm max}$.  Similarly, there is a change in the
nature of the fluctuations in the $y$-direction on passing through
$\phi_K$.  Below $\phi_K$ the fluctuations in $y$ are small but of
order~$\sigma$; above $\phi_K$, they will shrink at large $F$ like
those of the $x$-component, as~$1/F$.

Thus many properties of our system are similar to the properties of
real glasses.  One notable difference between our channel system and
the three dimensional hard sphere system is that in the latter system,
glass behavior occurs in the supercooled metastable region at
densities above that of the fluid-crystal transition.  Our own work is
a study of equilibrium behavior, in which the long-lived metastable
states are simply included in the thermal average.  But there are
parallels nonetheless between the two types of system above~$\phi_K$:
as the packing fraction increases towards its maximum value, both
systems develop a jammed crystal ordering.

One of our observations is that the structural feature which is behind
the onset of activated dynamics at $\phi_d$, which is
bond-orientational order in the channel problem, is not the same as
the structural feature which produces our $\phi_K$.  It is possible
that something similar may occur in three dimensional glasses, but
simulations of them at $\phi_K$ are difficult, so that the situation
remains unclear, like much else in the study of glasses.

\acknowledgments

We should like to thank Richard Bowles for supplying information on
the nature of the kink in the complexity $S_c(\phi)$.


\begin{thebibliography}{99}
\bibitem{quibble} $\Phi$ differs from the Gibbs free energy by an
  additive term that depends only on $N$ and $\beta$.  This is because
  the canonical partition function differs from the configuration
  integral by $2N$ factors of the thermal wavelength,
  $(2\pi\beta\hbar{}^2/m)^{1/2}$.
\bibitem{NNDelta} The expression for $\Delta_a$ is different in the
  nearest-neighbor case discussed in Ref.~\cite{Godfrey}.  For
  $h<\sqrt3\,\sigma/2$, removing a disk from one side of the channel
  leaves a gap of length $2\sqrt{\sigma^2-h^2}-\sigma$.  After this
  gap has been closed, the net increase in length per defect is
  $\Delta_a=\sigma - \sqrt{\sigma^2-h^2}$.
\end{thebibliography}
\end{document}